\documentclass[trackchanges,twocolumn]{aastex701}
\graphicspath{{./}{figures/}}
\usepackage{hyperref}

\usepackage{amsmath} 
\usepackage{tabularx}
\newcolumntype{Y}{>{\raggedright\arraybackslash}X}
\usepackage{graphicx}
\usepackage{subcaption}

\begin{document}

\title{Solar Vortices as Conduits for Magnetoacoustic Waves: Multi-Layer Coupling and Their Role in Atmospheric Heating}

\author[0000-0001-5414-0197,sname='Suzana S. A. Silva']{Suzana de Souza e Almeida Silva}
\altaffiliation{Plasma Dynamics Group, School of Electrical and Electronic Engineering}
\affiliation{Plasma Dynamics Group, School of Electrical and Electronic Engineering, University of Sheffield, Sheffield, S1 3JD, UK}\email[show]{suzana.silva@sheffield.ac.uk} 

\author[0009-0005-3577-4844,sname='Giannis Dakanalis']{Giannis Dakanalis}
\affiliation{Institute for Astronomy, Astrophysics, Space Applications and Remote Sensing, National Observatory of Athens, 15236 Penteli, Greece}\email[]{idakanalis@noa.gr} 

\author[0000-0001-5414-0197,sname='Luiz A. C. A. Schiavo ']{Luiz A. C. A. Schiavo}
\affiliation{Northumbria University, Newcastle upon Tyne, NE1 8ST, UK}\email{luiz.schiavo@northumbria.ac.uk} 

\author[orcid=0000-0002-2542-7073,sname='Tziotziou ']{Kostas~Tziotziou } \affiliation{Institute for Astronomy, Astrophysics, Space Applications and Remote Sensing, National Observatory of Athens, 15236 Penteli, Greece}\email{kostas@noa.gr}

\author[orcid=0000-0002-7711-5397]{Shahin Jafarzadeh}
\affiliation{Astrophysics Research Centre, School of Mathematics and Physics, Queen's University Belfast, Belfast, BT7 1NN, UK}\affiliation{Rosseland Centre for Solar Physics, University of Oslo, P.O. Box 1029, Blindern, 0315 Oslo, Norway}\email{shahin.jafarzadeh@qub.ac.uk}

\author[orcid=0000-0003-4747-4329]{Tiago M. D. Pereira}
\affiliation{Rosseland Centre for Solar Physics, University of Oslo, P.O. Box 1029, Blindern, 0315 Oslo, Norway}\affiliation{Institute of Theoretical Astrophysics, University of Oslo, P.O. Box 1029, Blindern, 0315 Oslo, Norway}\email{tiago.pereira@astro.uio.no}

\author[orcid= 0000-0002-8959-0806]{Georgia Tsiropoula}
\affiliation{Institute for Astronomy, Astrophysics, Space Applications and Remote Sensing, National Observatory of Athens, 15236 Penteli, Greece}\email{georgia@noa.gr}

\author[0000-0002-9546-2368]{Gary Verth}
\affiliation{Plasma Dynamics Group, School of Mathematical and Physical Sciences, University of Sheffield, Sheffield S3 7RH, UK}\email[]{g.verth@sheffield.ac.uk} 

\author[0000-0002-3066-7653]{Istvan Ballai}
\affiliation{Plasma Dynamics Group, School of Mathematical and Physical Sciences, University of Sheffield, Sheffield S3 7RH, UK}\email[]{i.ballai@sheffield.ac.uk}

\author[0000-0001-5597-1718]{Iñaki Esnaola}
\affiliation{Plasma Dynamics Group, School of Electrical and Electronic Engineering, University of Sheffield, Sheffield, S1 3JD, UK}\email[]{esnaola@sheffield.ac.uk}

\author[0000-0002-7863-624X]{James A. McLaughlin}
\affiliation{Northumbria University, Newcastle upon Tyne, NE1 8ST, UK} \email{james.a.mclaughlin@northumbria.ac.uk}

\author[0000-0002-5915-697X]{Gert J. J. Botha}
\affiliation{Northumbria University, Newcastle upon Tyne, NE1 8ST, UK}\email{gert.botha@northumbria.ac.uk}
            
\author[0000-0002-0893-7346]{Viktor Fedun}
\affiliation{Plasma Dynamics Group, School of Electrical and Electronic Engineering, University of Sheffield, Sheffield, S1 3JD, UK}\email[]{v.fedun@sheffield.ac.uk}


\begin{abstract}
Vortices are widespread throughout the solar atmosphere and commonly interpreted as channels for Alfv\'{e}n wave energy transport. However, this interpretation has mostly been inferred from the apparent alignment of rotating plasma structures with the magnetic field, rather than direct evidence of the nature of the waves involved. We present observational and numerical evidence that solar vortices act as coherent waveguides dominated by magnetoacoustic waves in the lower atmosphere, combining high-resolution SST observations and Bifrost simulation data. Using Normalised Mutual Information and Jensen-Shannon Divergence as coupling diagnostics, we provide the first observational identification of a solar vortex footpoint through information-theoretic metrics, locating it in the photospheric region of highest shared information with the chromospheric
swirl.  Spectral Proper Orthogonal Decomposition separates dominant dynamical patterns, allowing us to identify sausage, kink, and helical wave modes propagating along vortex structures. In the Bifrost simulation, we find no evidence for torsional Alfv\'en waves supported by vortices below 1 Mm. Instead, the waves supported by vortices are magnetoacoustic, and their energy transport is predominantly compressive in the lower atmosphere, with the magnetic component becoming dominant only at greater heights. Even the compressive wave component alone reaches nanoflare-level magnitudes ($\sim10^{25}$~erg), indicating that vortex-associated compressive perturbations may provide an energetically relevant contribution in regimes where magnetic-energy transport dominates. The recovery of sausage and helical modes in temperature perturbations shows that these modes carry a thermal signature, further supporting their relevance for chromospheric energy transport and localised heating.

\end{abstract}



\section{Introduction} 

The solar atmosphere hosts a wide range of vortical structures, including classical flow vortices driven by rotational motion (termed \textit{kinetic vortices}~\citep{Silva_2021}), \textit{magnetic vortices} involving twisted magnetic fields, and \textit{Poynting flux vortices}, where energy flows follow swirling trajectories~\citep{Silva_2024a}. Kinetic vortices alone occupy about 3\% of the photosphere~\citep{Giagkiozis_2018}, and occur in the chromosphere with a density of 10$^{-2}$~swirls~Mm$^{-2}$~min$^{-1}$, corresponding to a 2.65\% area coverage~\citep{Dakanalis_2021}.
These vortices are dynamic and multiscale, with diameters ranging from 500~km within intergranular lanes~\citep{Giagkiozis_2018} up to 5~Mm at supergranular flow junctions~\citep{Chian_2019, Tziotziou_2023}. They influence a wide spectrum of solar phenomena, from triggering flares in active regions~\citep{song2013} to modulating magnetic field dynamics and stability in the quiet Sun~\citep{Requerey_2017, Silva_2020, Yadav_2021, Battaglia_2021}. Complex systems such as \textit{solar tornadoes}~\citep{Tziotziou_2023} embody multiple vortex types, combining kinetic, magnetic, and energy vortices that transport electromagnetic energy upward~\citep{Silva_2024a}. Moreover, a vortex system enhances and concentrates electromagnetic energy \citep{Kuniyoshi_2023, Silva_2024a} and creates the necessary conditions for energy dissipation and plasma heating \citep{Yadav_2021, Silva_2024a, Silva_2024b}. Such vortex systems can connect the photosphere to the upper solar atmosphere \citep{Kitiashvili_2012, Wedemeyer2012, Silva_2020, Yadav_2021, Aljohani_2022}, manifesting themselves higher up as chromospheric swirls \citep{Tziotziou_2018,Shetye_2019,Dakanalis_2022} and creating dynamical energy channels \citep{Silva_2024a}. Thus, solar vortices are a key mechanism for explaining the local heating of the upper atmosphere and could be linked to observed phenomena such as campfires \citep{Kuniyoshi_2024}.

Given their influence on dynamics and stability \citep{Requerey_2017}, solar vortices are naturally expected  to act as effective waveguides, organising coherent perturbations over an extended vertical range. This guiding behaviour does not necessarily require a single magnetic flux tube to remain co-spatial with the vortex at all heights. Indeed, kink/helical and sausage-like disturbances can also arise in purely hydrodynamic vortical flows \citep{dahl2021centrifugal}, showing that the flow itself can organise wave-like perturbations. In the solar atmosphere, however, vortices evolve within a magnetised and strongly stratified plasma, where the magnetic field associated with a vortical structure may change connectivity or cease to coincide with the vortex at different heights \citep{yadav_2026}. The relevant waveguide is therefore better understood as a coupled plasma structure, produced by the combined organisation of the vortical flow, magnetic field, thermodynamic stratification, and transverse gradients. In this phenomenological sense, a vortex can provide a vertically coherent environment in which MHD perturbations are organised and transmitted through the atmosphere without requiring the vortex or its associated magnetic structure to preserve an identical geometry at every height.

Simulations incorporating vortical drivers at the footpoints of open magnetic flux tubes indeed excite guided MHD wave modes, including sausage, kink, and torsional Alfvén modes \citep{Fedun2011, Yadav_2020}. However, observational and simulation studies of spontaneously generated vortices (without imposed drivers) have predominantly reported incompressible wave excitation \citep{Jess_2009, Battaglia_2021}. Observationally, signatures of fast kink and standing wave modes have also been identified in chromospheric vortices \citep{Tziotziou_2019, Tziotziou_2020}. Hydrodynamic vortices support multiple wave modes, including kink and sausage modes, as shown by theory, simulations, and experiments \citep{Maxworthy_1988}. Their waveguiding capability depends on structural and dynamical properties, particularly vortex core thickness and interactions with boundaries or disturbances \citep{Maxworthy_Hopfinger_Redekopp_1985}. Flow vortices support helical waves, in which the vortex axis is displaced to produce a helical pattern. Such waves arise in swirling fluids with substantial axial flow and vorticity, as seen in atmospheric tornadoes, jet streams, and laboratory vortex experiments \citep{dahl2021centrifugal}.

In solar physics, studies of waves in vortices have largely focused on the excitation and transport of Alfvén waves \citep[e.g.][]{Wedemeyer2012, Battaglia_2021}. This focus rests on two implicit assumptions: first, that vortex motions necessarily generate twisted magnetic flux tubes (a condition not generally satisfied, particularly in the photosphere), see \cite{Silva_2021,Tziotziou_2023}); and second, that perturbations of such twisted fields lead to Alfvén pulses/waves \citep{Liu_2019, Battaglia_2021, Morton_2026}. 

In this study, we combine high-resolution SST/CRISP observations, realistic Bifrost simulations, and advanced analysis techniques to investigate the vertical coupling and wave dynamics of solar vortices. 
The wave modes were identified using Spectral Proper Orthogonal Decomposition SPOD, which is a data driven modal decomposition used for identify waves and coherent motion. SPOD is based in the classical 
Proper Orthogonal Decomposition (POD), which achieves a better separation of coherent motion \citep{sieber2016spectral}. POD was first introduced in solar physics for the detection of wave modes in the sunspots \citep{Albidah_2022, Albidah_2023} and pores \citep{Jafarzadeh_2024}. 
SPOD was first applied to identify large-scale compressible oscillations in the outer corona triggered by a coronal mass ejection \citep{Silva_2025}. More recently, SPOD was employed to identify wave modes in photospheric magnetic flux ropes \citep{Alanezy2026} and to characterize magnetoacoustic and Alfvén waves in numerical simulations of oscillatory  reconnection \citep{schiavo2026waves}.

This work is organized as follows, section 2 describes the observational, numerical, and synthetic datasets and the analysis methods employed. In Section 3, we quantify photosphere-chromosphere connectivity and identify the dominant MHD wave modes within vortices, followed by an assessment of their role in energy transport. Finally, Section 4 discusses the implications of our results for vortex-driven waveguiding of magnetoacoustic waves and chromospheric energetics.

\section{Data and Methodology} \label{sec:style}

\subsection{SPOD formulation\label{ref:mathSPOD}}
SPOD performs a modal decomposition of the fluctuating component of a spatiotemporal dataset $q(\mathbf{x},t)$. The dataset is first separated into a time-averaged component and a perturbation field according to
\begin{equation}
  q(\mathbf{x},t) = \langle q(\mathbf{x}) \rangle + q^\prime(\mathbf{x},t),
  \label{eq:pert}  
\end{equation}
where $\langle q(\mathbf{x}) \rangle$ denotes the temporal mean of the dataset and $q^\prime(\mathbf{x},t)$ represents the fluctuations about this mean state. The perturbation field is then expressed as a superposition of SPOD modes, 
\begin{equation}
q^\prime(\mathbf{x},t) = \sum_{n=1}^{N} \phi^{(n)}(\mathbf{x})\, a^{(n)}(t),
\end{equation}
where the superscript $n$ indexes the mode, and $N$ is the number of temporal samples (snapshots) used in the decomposition. The spatial modes and temporal coefficients are given by
\begin{equation}
\phi^{(n)}(\mathbf{x}) = \frac{1}{\lambda_n N} \sum_{k=1}^{N} \xi_{k,n}\,  q^{\prime}\left(\mathbf{x},t_k\right)
\mbox{,}
\end{equation}
and
\begin{equation}
a^{(n)}(t_k) = \sqrt{N\lambda_n}\, \xi_{k,n} \mbox{,}
\end{equation}
where $\lambda_n$ and $\boldsymbol{\xi}^{(n)}$ are the eigenvalue and eigenvector associated with the $n$-th SPOD mode, and $\xi_{k,n}$ is the $k$-th component of $\boldsymbol{\xi}^{(n)}$.  

Together, the spatial modes and temporal coefficients provide a low-dimensional representation of the dominant coherent structures and their temporal evolution within the dataset.
They are obtained from the eigenvalue problem
\begin{equation}
Z \boldsymbol{\xi}^{(n)} = \lambda_n \boldsymbol{\xi}^{(n)} \mbox{,}
\end{equation}
where $Z$ is a filtered covariance matrix defined as 
\begin{equation}
Z_{i,j} \;=\; \sum_{k=-N_f}^{N_f} g_k\, C_{i+k,\, j+k} \mbox{ ,}
\end{equation}
where $N_f$ is the filter half-width and $g_k$ are the coefficients of a symmetric kernel. In this work, $g_k$ is chosen as a Gaussian kernel, following \cite{Ribeiro_2017}. The (unfiltered) snapshot covariance matrix, $C$, is
\begin{equation}
C_{t_{i},t_{j}} \; = \; \frac{1}{N} \int_{\Omega} q^{\prime}\left(\mathbf{x},t_{j}  \right)  
q^{\prime} \left(\mathbf{x},t_{i}  \right) d\mathbf{x} \mbox{,}
\label{eq:covmatrix}
\end{equation}
where $C_{t_i,t_j}$ is the covariance between the fluctuation fields at snapshots $t_i$ and $t_j$, and $\Omega$ denotes the spatial domain over which the inner product is evaluated. 
SPOD differs from classical POD in that a filter is applied along the diagonals of $C$ prior to the eigendecomposition,

For the analysis of the 3D MHD simulations, a vectorial formulation of SPOD was employed using magnetic-energy norm. This approach allows the decomposition to isolate coherent wave-mode structures according to their contribution to the magnetic-energy dynamics of the system, enabling the identification of the dominant modes governing magnetic-energy transport throughout the simulated atmosphere. In this case, Eq.~\ref{eq:covmatrix} becomes

\begin{equation}
C_{t_i,t_j} = \frac{1}{N} \int_{\Omega}
E_{\mathrm{mag}}^{\prime}(\mathbf{x}, t_i)
E_{\mathrm{mag}}^{\prime}(\mathbf{x}, t_j)
\, d\mathbf{x},
\label{eq:vec-SPOD}
\end{equation}

\noindent where $E_{\mathrm{mag}}^{\prime}$ represents the magnetic-energy fluctuations, defined analogously to Eq.~\ref{eq:pert} as deviations from the temporal mean magnetic energy. The temporally filtered covariance matrix derived from $E_{\mathrm{mag}}^{\prime}$ was subsequently used to project and decompose the remaining physical variables, ensuring that the extracted modes correspond to coherent structures ranked according to their magnetic-energy content.

By filtering the covariance matrix, the frequencies of each mode are constrained, and the modes will mostly display only one frequency. An interesting aspect of this filtering step is that it allows the ranking of SPOD modes, i.e., the modes are ranked by their contribution to the perturbations of the original data/signal.

\subsection{Automated detection of chromospheric swirls with the A-MorphIS code}
\label{sec:amorphis}

The \textbf{A}utomated \textbf{Morph}ological \textbf{I}dentification of \textbf{S}wirls (A-MorphIS) code \citep{Dakanalis_2021} has been used for the automated detection of swirls within the field-of-view (FOV) of the chromospheric observations datacube. Given the challenges in deriving reliable horizontal velocity vector fields in the highly dynamic chromosphere with Local Correlation Tracking (LCT) detection techniques  \citep{Tziotziou_2023}, A-MorphIS instead utilises the morphological features of swirling structures. In chromospheric spectral lines, such as the line-center and near-line-center wavelengths of H$\alpha$ and Ca\,{\sc ii} IR, these structures manifest as rapidly rotating absorption patterns (e.g., dark spirals, circles, arcs). However, while A-MorphIS has been initially tailored for chromospheric data, it can be applied to any datacube, observational or simulated, that contains absorption-like features indicative of swirling motions.

The detection approach involves several distinct and sequential stages that lead to the identification of high-point density areas of traced segments of curved chromospheric features. The approach draws on techniques such as curvature center density, and winding angle methods, which are broadly used in deriving streamline velocity fields across various scientific disciplines. The process begins by tracing dark circular and spiral-like features within each high-contrast, edge-enhanced frame of the temporal data cube,  constructed through appropriate filtering and adaptive (local) thresholding techniques. Following this, a minimum radius threshold and a set of angle-related curvature criteria are applied to retain only the segments exhibiting high curvature along with their corresponding centers of curvature. An unsupervised machine learning technique is then employed to cluster these remaining centers of curvature within each frame, identifying regions of high-point density which are labelled as swirl candidates (SCs). In a final step, the SCs from each frame are projected into a common two-dimensional space, namely the FOV of the observations. A second-level clustering is then applied, revealing new high-point-density areas in the temporal dimension, thereby determining the final swirl locations and their temporal evolution. Additionally, the radius of the detected swirls is computed using the outermost 10\% of traced swirl segments.

The algorithm has already been tested in H$\alpha$  CRISP/SST observations \citep{Dakanalis_2022}, uncovering an unprecedented population of swirling events in the chromosphere. This has enabled a detailed statistical analysis of their characteristics, including lifetime, radius and intermittency.

\subsection{Observations}
The multi-line, multi-wavelength datasets, including the H$\alpha$ and the Ca\,{\sc ii}\,8542\,\AA\ (hereafter Ca\,{\sc ii}) spectral lines used in our analysis, were acquired on August 13, 2019 with the CRisp Imaging Spectro-Polarimeter \citep{Scharmer_2008} (CRISP) mounted on the Swedish 1-m Solar Telescope \citep{Scharmer_2003} (SST). The observations targeted a quiet-Sun region at the solar disk center, covering a FOV of 57$\times$58 arcsec$^2$ with a pixel size of 0.0592 arcsec. For H$\alpha$, data were collected at thirteen equally spaced wavelengths across the spectral window [-1.2 \AA, 1.2 \AA] from line-center (6563\,\AA) with a 0.2 \AA\ spectral step and a narrowband filter of 0.061 \AA. Ca\,{\sc ii}  observations were taken at eleven wavelengths (line center, $\pm$0.1 \AA, $\pm$0.2 \AA, $\pm$0.3 \AA, $\pm$0.4 \AA, $\pm$0.6 \AA) using a filter with a bandwidth of 0.107 \AA. 

Both spectral lines were recorded simultaneously from 10:20:37 UT to 10:54:24 UT with an average cadence of $\sim$6.3~s. The raw data were processed with the SSTRED \citep{Lofdahl_2021} reduction pipeline, which includes image restoration through Multi-Object Multi-Frame Blind Deconvolution \citep{vannoort_2005} (MOMFBD) to correct for atmospheric distortion. After removing 16 highly-blurred images during processing, each spectral line resulted in a data cube of 304 images. A minor displacement in the telescope pointing about 25~min into the observations required a new co-alignment of the images, resulting in a common rectangular FOV of 51.3$\times$53.6 arcsec$^2$ for the final data cubes.  

\begin{figure*}[htp!]
    \centering
    
    \includegraphics[width =\textwidth]{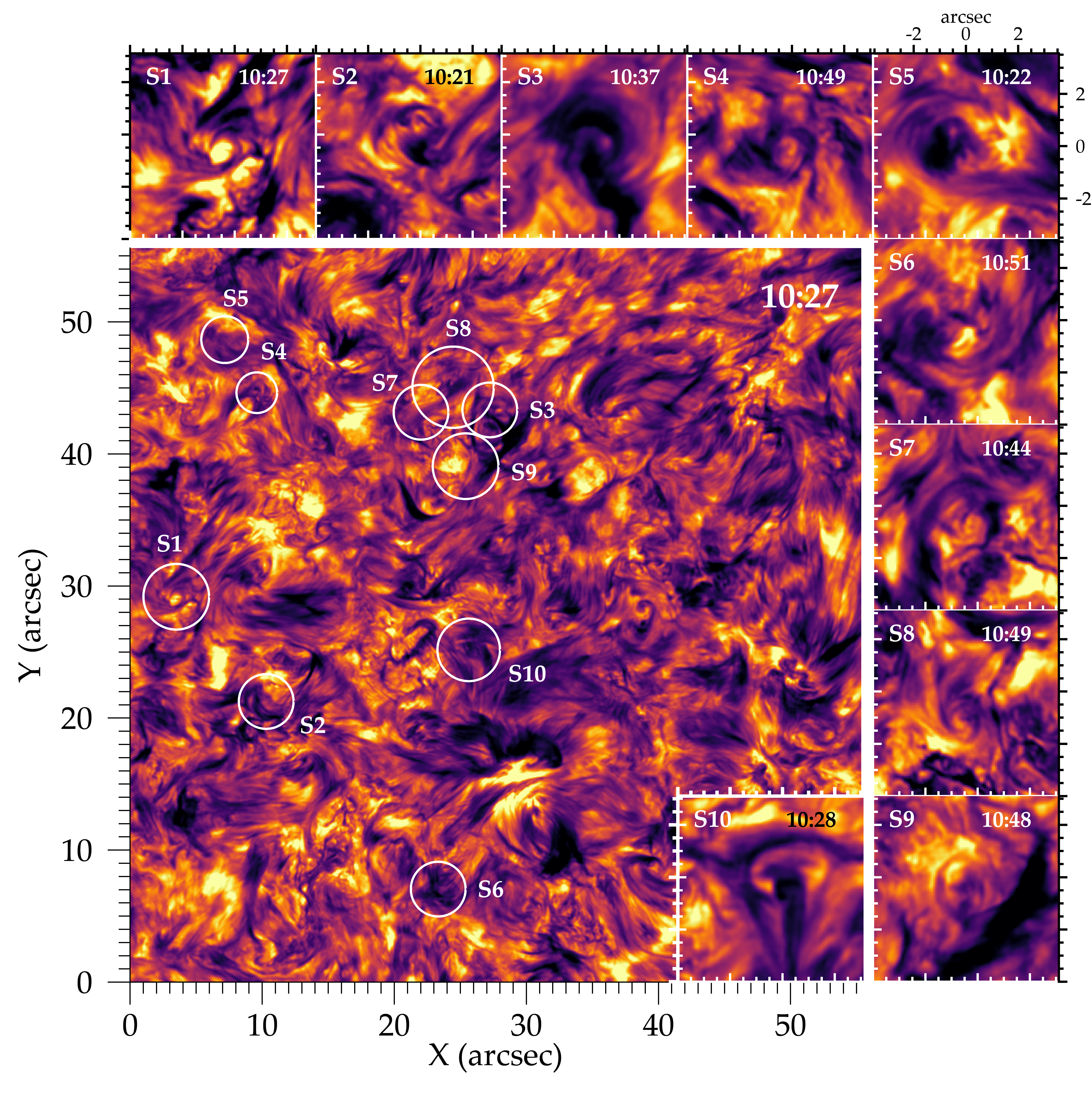}
    
    \caption{Snapshots of the 10 analysed swirl events in H$\alpha$ listed in Table~\ref{tab:my_label}, along with snapshot 65 showing the full FOV. Each panel in the top row and right column displays a 6$\times$6 arcsec subregion centered on an individual swirl, cropped from the full H$\alpha$ FOV to highlight the characteristic morphology of each swirling structure. The selected frames represent high-quality moments in the temporal evolution of the swirls. The central panel shows the full H$\alpha$ FOV, with swirl locations marked by circles centred at the mean positions and radii computed using the A-MorphIS detection algorithm.}
    \label{fig:fov}
\end{figure*}

Figure~\ref{fig:fov} presents snapshots of the analysed FOV, which is populated with numerous persistent swirls observed in both spectral lines, that appear, however, more prominently in H$\alpha$. Table~\ref{tab:my_label} summarises the physical characteristics in H$\alpha$ and provides visual details for ten chromospheric swirls identified in our observations using the A-MorphIS detection code and used for our analysis. The locations of these swirls are clearly marked in Fig.~\ref{fig:fov} 
and \cite{Dakanalis_2022} also show a movie of that observation.

\begin{deluxetable*}{r c c l}[ht]
\digitalasset
\tablewidth{0pt}
\tablecaption{Physical properties in H$\alpha$ and visual characteristics in both H$\alpha$ and Ca\,{\sc II} for the 10 observed swirls selected for the analysis. The swirls were identified in H$\alpha$ filtergrams from the CRISP dataset using the automated A-MorphIS detection algorithm. Swirl radii were determined based on the outermost 10\% of the traced segments, as described in the text. \label{tab:my_label}}
\tablehead{
\colhead{Swirl} & \colhead{duration (min)} & \colhead{diameter (arcsec)} & \colhead{Visual characteristics}
}
\newcommand{\vcell}[1]{\parbox[t]{0.55\textwidth}{\small #1}}
\startdata
       S1  & 10   & 1.5  & \vcell{Persistent presence in both H$\alpha$ and Ca\,{\sc II}, exhibiting a circular or spiral-like shape, and associated with a photospheric magnetic bright point (MBP) located offset from the swirl center} \\
S2  & 19   & 1.52 & \vcell{Visible for nearly the entire duration, more prominent in H$\alpha$ than in Ca\,{\sc II}, and associated with a single MBP directly beneath the swirl center} \\
S3  & 33.7 & 2.25 & \vcell{Visible for almost the entire duration in both H$\alpha$ and Ca\,{\sc II}, exhibiting numerous displacements and shape changes, and associated with multiple photospheric MBPs that rearrange over time} \\
S4  & 4.4  & 1.8  & \vcell{Persistent presence, more prominent in H$\alpha$, and associated with a shape-shifting photospheric MBP} \\
S5  & 6.7  & 1.7  & \vcell{Persistent presence, more prominent in H$\alpha$, associated with a photospheric MBP located slightly off-center from the swirl} \\
S6  & 6.4  & 1.8  & \vcell{Persistent presence in both H$\alpha$ and Ca\,{\sc II}, exhibiting circular and spiral-like shapes, with no visible photospheric MBP near the swirl center} \\
S7  & 6.5  & 1.5  & \vcell{Persistent presence in H$\alpha$, limited visibility in Ca\,{\sc II}, with no visible photospheric MBP near the swirl center, but multiple MBPs present within the swirling area}  \\
S8  & 6.6  & 1.5  & \vcell{Persistent presence in both H$\alpha$ and Ca\,{\sc II}, associated with a single, shape-shifting photospheric MBP directly beneath the swirl center} \\
S9  & 8.7  & 1.1  & \vcell{Persistent presence in both H$\alpha$ and Ca\,{\sc II}, exhibiting circular and spiral-like shapes, with a visible, static photospheric MBP directly beneath the swirl center} \\
S10 & 8.7  & 1.3  & \vcell{Persistent presence in H$\alpha$, moderate visibility in Ca\,{\sc II}, exhibiting circular and spiral-like shapes, and appears to be dragging material from the surrounding area, with no visible photospheric MBP near the swirl center} \\
\enddata
\end{deluxetable*}

\subsection{Bifrost Simulation and synthetic spectra}
We analyse the Bifrost simulation ch024031\_by200bz005, which was developed within the Hinode Science Data Centre framework for direct comparison with high-resolution observations of the solar atmosphere \citep{Carlsson2016, Finley_2022}.
The simulation domain consists of a 768$\times$768$\times$768 grid, spanning 24 Mm in both horizontal directions and 16.8 Mm in the vertical direction. The horizontal resolution is 31 km, while the vertical resolution is non-uniform, ranging from 12 to 82 km. The model includes realistic radiative transfer, magnetic field evolution, and its atmosphere extends into the lower corona, reaching up to 14.3 Mm above the photosphere. It represents a coronal hole environment, characterised by predominantly open magnetic field lines in the upper atmosphere and a photospheric field composed of small-scale mixed polarities, with an average unsigned field strength of 40 G.   More details on this particular dataset can be found in \cite{Finley_2022}. The use of a coronal-hole setup is an advantage, as the simpler magnetic topology reduces large-scale magnetic interference. This allows vortex-driven dynamics and their associated coupling signatures to be isolated more clearly, making this dataset ideal for our study,  which focuses on waveguidance by vortices in the lower atmosphere.

\begin{figure*}[htp!]
    \centering
    \includegraphics[trim={0mm 50mm 0mm 20mm}, clip, width =1\textwidth]{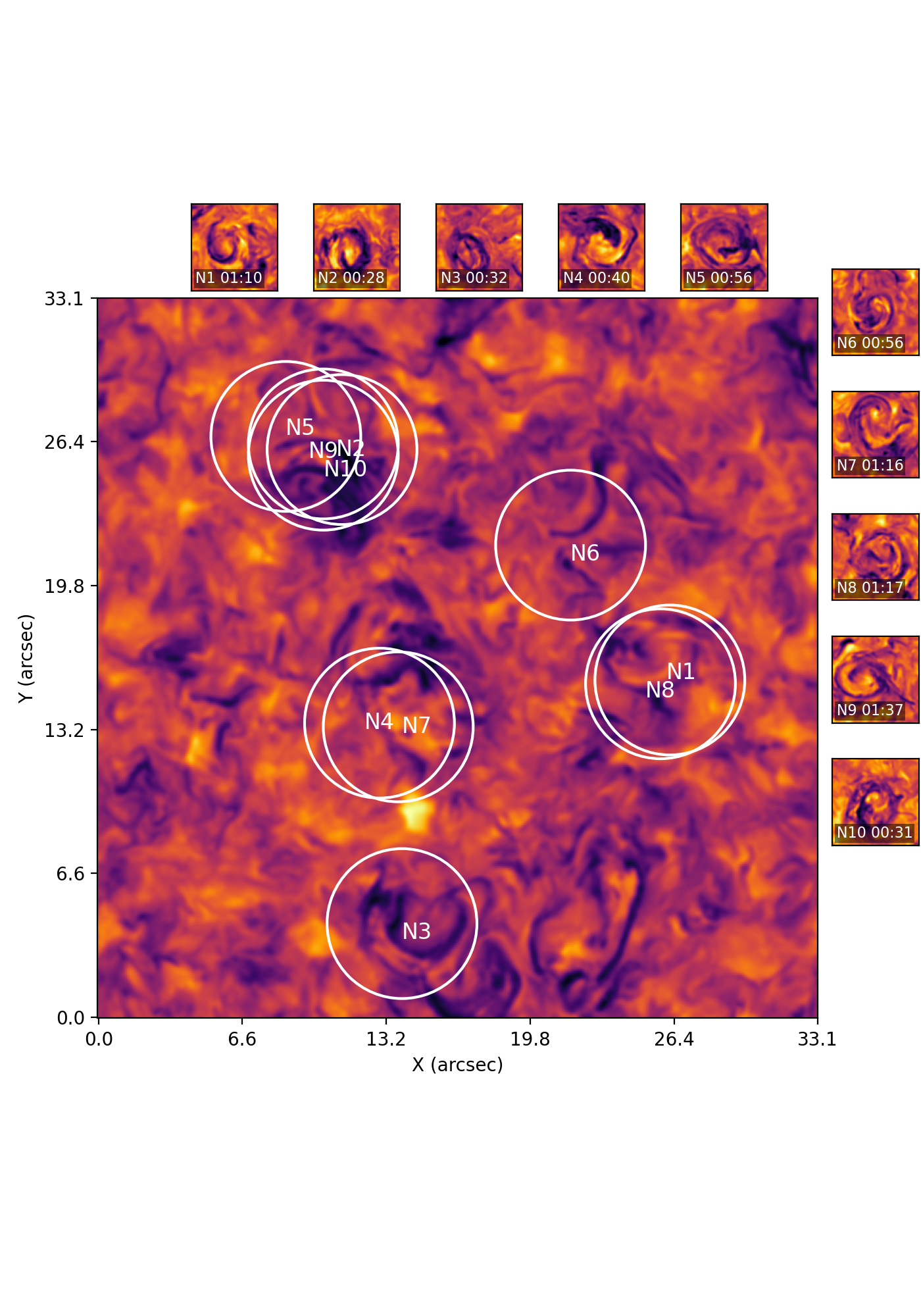}
    \caption{Selected swirls for analysis in the Bifrost simulation. The main image shows the H$\alpha$ core at $t=1$ h and 31 min and the swirl regions are depicted by white circles. The subplots show a close view of the selected swirls.}
    \label{fig:useddomains}
\end{figure*}

For wave mode analysis, we used time series of synthetic H$\alpha$ spectra. The selected swirls for analysis are shown in Fig.~\ref{fig:useddomains}. They were identified using the $\Gamma$-method \citep{graftieaux} on the velocity field. The spectra were computed allowing for departures from local thermodynamic equilibrium (non-LTE). To reduce computational costs, we used SunnyNet \citep{Chappell2022}, which uses convolutional neural networks to approximate non-LTE hydrogen populations. SunnyNet was trained on two simulation snapshots run with Multi3D  \citep{Leenaarts_2009}. Based on the SunnyNet populations, the final spectra were obtained using the Muspel library  \citep{Pereira_2025}.

\subsection{Synthetic data model}
In order to validate and interpret our findings, we generated synthetic data that mimics the cross-section of a vortex tube in both the photosphere and the chromosphere as an elliptical Gaussian-like distribution. While direct observational or numerical studies of vortex deformation under flow strain in solar MHD are lacking, this choice is physically motivated by classical fluid dynamics, where vortices, subject to external strain, often maintain elliptical shapes \citep{kida1981elliptical}. The center of the distribution, $(x_c, y_c)$, was allowed to undergo a combination of random walk and oscillatory motion, mimicking the transverse displacement of the vortex structure. This behaviour is consistent with the signature of kink modes, which involve collective, transverse displacements of magnetic flux tubes or vortical structures. The coordinates $(x_c, y_c)$ are given by
\begin{align}
	x_c &= x_c^{\text{walk}} + A_{\text{kink}} \cos(\omega_{\text{kink}} t), \\
	x_c^{\text{walk}} &\leftarrow x_c^{\text{walk}} + \mathcal{S}\,\xi_x, \\
	y_c &= y_c^{\text{walk}}, \\
	y_c^{\text{walk}} &\leftarrow y_c^{\text{walk}} + \mathcal{S}\,\xi_y,
\end{align}
where \(A_{\text{kink}}\) is the amplitude of the transverse kink displacement, 
\(\mathcal{S}\) is the scale factor controlling the random step size, and 
\(\xi_x\) and \(\xi_y\) are random variables independently sampled at each timestep from a uniform distribution
\begin{align}
	\xi_x,\xi_y \sim \mathcal{U}(-A_{\text{walk}}, A_{\text{walk}}).
\end{align}
Based on the observed dynamics of SST swirls, we chose 
\(A_{\text{walk}}=0.03\) pixels and \(\mathcal{S}=0.0764\).

The sausage mode is mimicked by modulating the width of the Gaussian profile representing the vortex. Specifically, we define the time-dependent width parameter, \( \sigma_v \), as:
\begin{equation}
	\sigma_v(t) = 1 + A_{\text{S}} \cos(\omega_{\text{S}} t),
\end{equation}
where \( A_{\text{S}} \) and \( \omega_{\text{S}} \) denote the amplitude and frequency of the sausage mode, respectively.

Thus, the synthetic intensity distribution, \(G(x,y,t)\), representing a vortex waveguide supporting both kink and sausage modes, is given by
\begin{equation}
	G(x, y, t) = \exp\left( -\frac{s_x (x' - x_c)^2 + s_y (y' - y_c)^2}{2 \sigma_v(t)^2} \right),
\end{equation}
where \( (x_c, y_c) \) is the time-dependent center of the vortex defined earlier and represents the kink-mode displacement, and \( s_x \) and \( s_y \) are ellipticity parameters that control the stretching of the Gaussian in the \( x \) and \( y \) directions, respectively.
The primed coordinates \((x',y')\) are obtained by rotating the original spatial coordinates \((x,y)\) around the vortex center to simulate the torsional motion of the vortex
\begin{align}
	x' &= \cos(\omega_{\rm rotation} d t)\,x - \sin(\omega_{\rm rotation} d t)\,y, \\
	y' &= \sin(\omega_{\rm rotation} d t)\,x + \cos(\omega_{\rm rotation} d t)\,y,
\end{align}
where \(\omega_{\rm rotation}\) is the angular rotation frequency and \(d\) is a dimensionless parameter controlling the radial dependence of the angular velocity. For solid-body rotation we set \(d=1\). For differential rotation, we define
\begin{equation}
	d = 0.25 \frac{\sqrt{(x-x_c)^2+(y-y_c)^2}}{\sigma_v},
\end{equation}
such that the angular velocity varies linearly with the distance from the vortex center. The normalization by \(\sigma_v\) ensures that \(d\) remains dimensionless.

To account for the observational noise commonly present in SST data, we added a random noise field sampled independently at each pixel from a uniform distribution
\begin{align}
	\mathrm{noise}(x,y) \sim \mathcal{U}(0,0.2),
\end{align}
where \(\mathcal{U}(a,b)\) denotes a uniform random distribution in the interval \([a,b]\). The Gaussian distribution obtained in this way evolves over discrete time steps \(t_n\), with \(t \in [0,60]\)  such that the synthetic data have the same cadence as the SST observations. At each timestep, the field \(G(x,y,t)\) is updated using the parameters described above.

The centers of these distributions undergo a combination of random walk and a kink oscillatory motion in the $x$-direction, while their widths evolve sinusoidally to emulate sausage modes. Vortex motion is incorporated with two models considered: radially varying angular velocity (model I, differential rotation) and solid-body rotation (model II). Spatial scales and time cadence were chosen to match observational data (CRISP/SST). The models and their full SPOD analysis are presented in Appendix \ref{app:spod-model}

\subsection{Information Theory}

\subsubsection{Mutual Information}
Mutual Information (MI) \citep{cover2012elements} quantifies the dependence between two random variables through the reduction of uncertainty each provides about the other. It measures the amount of shared information, is symmetric and non-negative, and equals zero only when the variables are completely independent.  
In this study, MI was computed between five photospheric regions and a fixed chromospheric region centred on the swirl, in the same FOV, to assess their statistical dependency. 
The expression of MI is given by
\begin{equation}
\mathrm{MI}(X;Y) = \sum_{x\in\mathcal{X},y\in\mathcal{Y}} p(x,y) \log \left( \frac{p(x,y)}{p(x)p(y)} \right),
\end{equation}
where, \(\log\) denotes the natural logarithm, such that the mutual information is expressed in nats, $p(x, y)$ is the joint probability distribution of variables $X$ and $Y$ taking values in $\mathcal{X}$ and $\mathcal{Y}$, respectively, and $p(x)$ and $p(y)$ are their respective marginal probability distributions. High MI between two atmospheric layers indicates coupling or shared dynamics, whereas low MI signifies statistical independence or weak interaction.

The MI computation was performed using a discrete histogram-binning approach applied to spatially averaged time series. Each data cube was first standardised using $Z$-score normalisation, transforming each pixel's time series to zero mean and unit variance:
\begin{equation}
X_\mathrm{norm} = \frac{X - \mu}{\sigma + \epsilon},
\label{eq:zscore}
\end{equation}
where $\mu$ and $\sigma$ are the mean and standard deviation computed over time for each pixel, and $\epsilon = 10^{-8}$ ensures numerical stability. Each standardised data cube was then reduced to a single representative time series by computing the spatial mean over the region at each time step. Normalisation removes absolute intensity biases between regions and instruments, so that the MI reflects shared temporal patterns rather than differences in scale or baseline offset.

Per-series bin edges were constructed independently for each spatially averaged time series, with the number of bins set to $\max(3,\lfloor N_t/5 \rfloor)$, where $N_t$ is the number of time steps. The time series were discretised with \texttt{numpy\.digitize} using per-series bin edges, and MI was computed from the resulting bin indices using \texttt{mutual\_info\_score} from \texttt{scikit-learn}. The Shannon entropies $H_X$ and $H_Y$ were computed from the same discretised bin counts as $H = -\sum_k p_k \ln p_k$ \citep{cover2012elements}, ensuring unit consistency with the MI estimate. The normalised mutual information is then defined as
\begin{equation}
\mathrm{NMI} = \frac{\mathrm{MI}(X;Y)}{\min(H_X,\, H_Y)},
\end{equation}
which lies in the interval $[0,1]$. An NMI  value of 0.5 indicates that 50\% of the target variable's entropy can be explained by the source, which represents a qualitative description of coupling strength.

\subsubsection{Jensen-Shannon Divergence}

Jensen-Shannon Divergence (JSD) \citep{cover2012elements} is a symmetric measure that quantifies the similarity between two probability distributions. The JSD between two distributions, $P$ and $Q$, is defined as
\begin{equation}
\mathrm{JSD}(P \parallel Q) = \frac{1}{2} KL(P \parallel M) + \frac{1}{2} KL(Q \parallel M), 
\end{equation}
where $M = \frac{1}{2}(P + Q)$ and $KL(P \parallel M)$ denote the Kullback-Leibler divergence between $P$ and the average distribution $M$. Physically, JSD quantifies how different the distributions of two datasets are. Low JSD between photospheric and chromospheric regions indicates similar value distributions and potentially synchronised dynamics, whereas high JSD implies dissimilar, more independent behaviour. JSD, therefore, complements NMI by highlighting statistical divergence, while NMI quantifies dependency or shared information.

Before computing JSD, both data cubes were normalised jointly to the $[0,1]$ range using global min-max normalisation:
\begin{equation}
x_\mathrm{norm} = \frac{x - x_\mathrm{min}}{x_\mathrm{max} - x_\mathrm{min} + \epsilon},
\end{equation}
where $x_\mathrm{min}$ and $x_\mathrm{max}$ are taken over the combined distribution of both cubes, ensuring histogram bins fall within a common numerical range and preventing magnitude-related distortions. JSD was computed from normalised histograms of the flattened data cubes using a base-2 logarithm, yielding values in $[0,1]$.

\section{Results}
\subsection{Vortex photosphere - chromosphere connectivity}

\begin{figure}
\centering
    \includegraphics[width =0.5\textwidth]{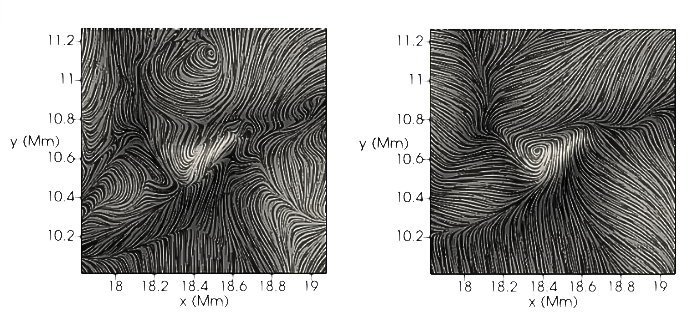}
    \caption{Here, the synthetic H$\alpha$ wing is overlaid with the Line Integration Convolution (LIC) \citep{cabral_93} of the magnetic (left panel) and velocity fields (right panel) computed at 100 km above the solar surface. Darker LIC lines indicate vectors tangent to the field along the plane. 
    }
    \label{fig:useddomain}
\end{figure}

\begin{figure*}
    \includegraphics[width=1.0\textwidth]{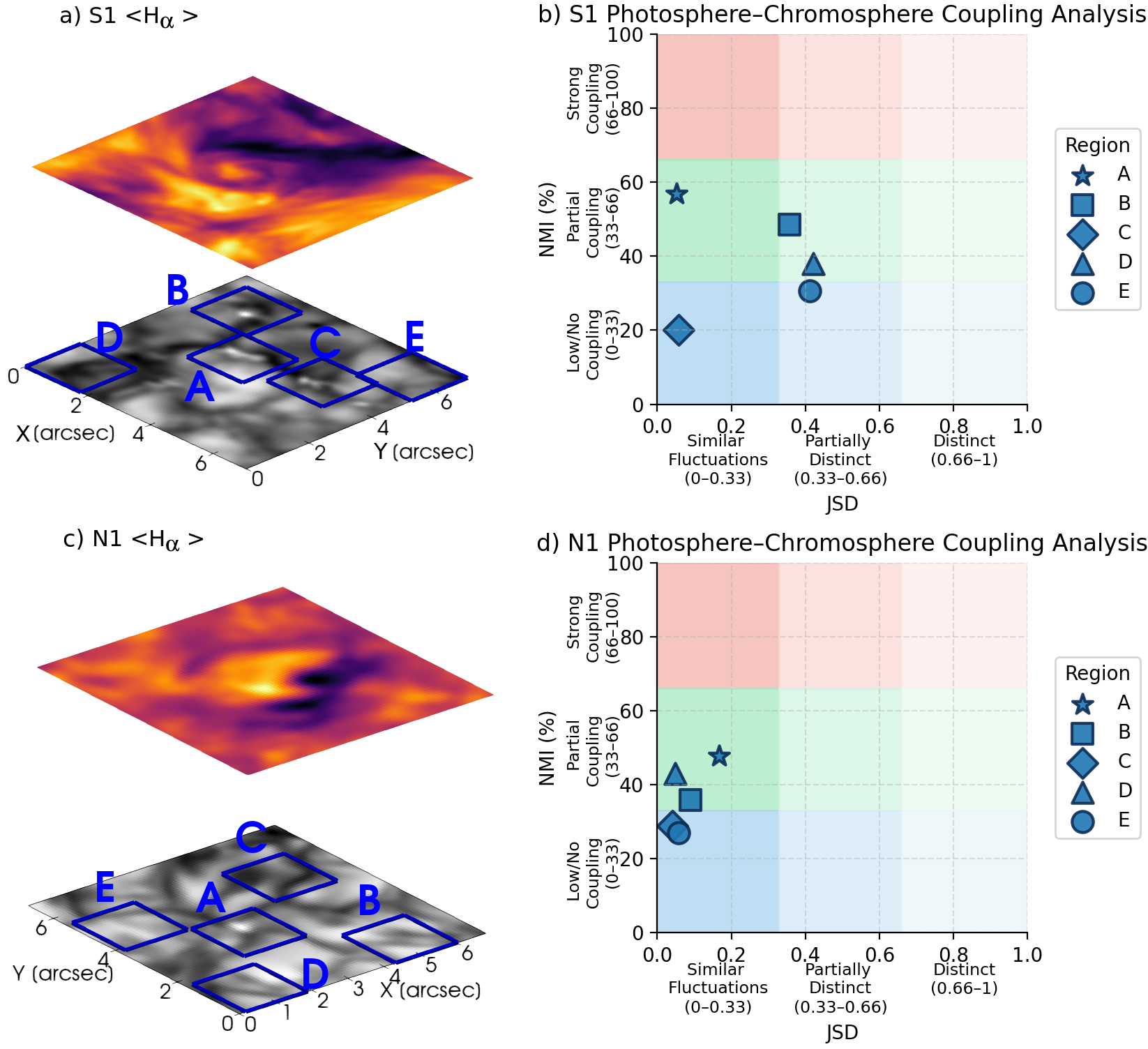}
    \caption{Photosphere - chromosphere coupling analysis using Shannon entropy and Jensen-Shannon divergence (JSD). Panels (a - b) correspond to the observational vortex (S1), while panels (c - d) refer to the numerical vortex simulation (N1). (a, c) Close view of H$\alpha$ time average intensity maps for selected vortices. The time interval used for the averaging and NMI/JSD analysis is the same as used in the wave identification section. The photospheric base, H$\alpha$ wing, presents labelled sub-regions A to E used for NMI and JSD analysis. (b, d) Coupling analysis based on JSD ($x$-axis) versus NMI ($y$-axis) across the selected sub-regions, quantifying the degree of similarity in the spatio-temporal structures between the two layers. The background colour highlights the coupling regimes: low coupled (blue, NMI $< 33$), partially and moderate coupling (green, $33 \leq \mathrm{NMI} < 66$), and strong coupling (red, $\mathrm{NMI} \geq 66$). Symbol shapes denote corresponding sub-regions as labelled in panels (a) and (c).
}
\label{fig:connectivity}
\end{figure*}

We selected two representative vortices labelled S1 (from SST observational data, Fig.~ \ref{fig:fov}) and N1 (from numerical simulation, Fig.~ \ref{fig:useddomains}). A close view of those vortices are shown in Fig. \ref{fig:connectivity}.
In the simulation, prior analysis shows that N1 is a 3D vortex that extends from the photosphere to the upper atmosphere \citep{Silva_2024a}. In the H$\alpha$ core, N1 presents the expected chromospheric swirl signature as shown both in Figs. \ref{fig:useddomains} and \ref{fig:connectivity}. The photospheric vortex footpoint of N1 is located in region A, and it appears as a bright point in the  H$\alpha$ wing. To use the same metrics for both numerical and observational data, our vortex identification technique should not depend on velocity fields, as it is currently not possible to retrieve such flow properties from chromospheric observational data. 
Thus, the pattern in the H$\alpha$ core was used to determine the vortex region and general shape of the vortex system in the chromosphere, as done by other studies \citep{Shetye_2019, Dakanalis_2021, Dakanalis_2022}. 
In the photospheric counterpart, a vortex can be delimited by the bright H$\alpha$ wing, as indicated in
Fig.~\ref{fig:useddomain} for the simulation vortex N1.

Figure~\ref{fig:connectivity} (left panels a and c) presents a close-up view of the vortex regions. The chromospheric cross section is indicated by the H$\alpha$ core map on the top, while the photospheric counterpart of the vortex is represented by the H$\alpha$ wing surface on the bottom slice. 
In the observation, S1 clearly presents the chromospheric swirl signature for the H alpha core line, and it has been previously analyzed by \cite{Dakanalis_2021, Dakanalis_2022}. Obtaining the photospheric vortex footpoint in observations is not straightforward, as at the wing wavelength, we have three distinct bright points within the field of view. Thus, we needed a way to define which bright point corresponds to the actual vortex footpoint. 

To identify the photospheric footpoint of S1 swirl among the brightpoints, we employ an information-theoretic approach based on NMI and JSD. These metrics offer a principled, model-free framework for quantifying the statistical relationship between intensity distributions at two atmospheric layers, without assuming any particular functional form for the coupling between them.
If a photospheric vortex and a chromospheric swirl encompass the same magnetic flux tube forming a 3D vortex tube (as established from simulations \citep{Wedemeyer2012, Silva_2024a}), the vortex-driven perturbations propagating along the vortex tube are expected to imprint a common statistical signature on both layers. This shared dynamical history should manifest as a high degree of shared information (reflected in elevated NMI values) together with a relatively low divergence between the respective fluctuation distributions (reflected in low JSD values). Conversely, spatially or magnetically unrelated regions are expected to exhibit statistically independent fluctuations, yielding low NMI and high JSD.

We computed NMI/JSD values between selected photosphere regions and the chromosphere swirl for both N1 and S1.  For our calculation, we restricted the time window to the same interval used for the wave analysis. 
This ensures that the vortex-breakdown phase is excluded from the analysis, since the associated loss of coherent rotational structure could obscure coupling signatures and bias the resulting estimates \citep{LUCCANEGRO2001431}.

For each region of interest (A, B, C, D, and E), our calculation was performed comprising 20$\times$20 pixels for N1 and 30$\times$30 pixels for S1. 
The region size was determined by the spatial extent of the bright-point-containing structures in each case: Region A for simulated vortex N1, and Regions A, B, and C for observed swirl S1. Each region was defined to be as small as possible while fully encompassing the spatial motion of the bright points throughout the analysed time interval. 

Instead of computing NMI/JSD based on intensity, which can be more influenced by background variations, our calculations use intensity perturbations. This isolates deviations from equilibrium, which are the physically meaningful quantities in the context of vortex-induced coupling between the photosphere and chromosphere.

For each dataset I, we defined perturbation cubes as
\[
I_{\mathrm{pert}}(t,y,x) = I(t,y,x) - \left\langle I(y,x) \right\rangle_t ,
\]
where \(\left\langle \cdot \right\rangle_t\) denotes the temporal average.  Here, the intensity cube is written in array-index notation as $I(t,y,x)$, where $t$ denotes time and $(x,y)$ the spatial coordinates. The photospheric perturbation cube for each region was then spatially scaled up to the chromospheric region size before the entropy-based diagnostics were computed.

The results of the NMI/JSD analysis are presented in the second column of Fig.~\ref{fig:connectivity}, where the $y$-axis corresponds to NMI values and the $x$-axis to JSD values for each subregion. In a magnetised solar atmosphere, global p-mode oscillations and the continual evolution of the magnetic network can introduce background correlations across atmospheric layers. Consequently, all analysed photospheric subregions in both the numerical and observational datasets exhibit non-zero NMI with their corresponding chromospheric swirl, indicating some degree of shared dynamical information. Likewise, the perturbation distributions are not fully dissimilar, as reflected by JSD values below 0.6.

 For the simulation (panel d), the vortex N1 footpoint (Region A) exhibits the highest NMI value, corresponding to 47.8\% shared information, while the control regions span 27.0\% to 43.0\%. Region A thus exceeds the control values by 4.9 to 20.8 percentage points, corresponding to a relative enhancement of approximately 11.3\% to 76.9\% over the control regions. The non-footpoint regions (B to E) show a narrow, uniformly low JSD range ($<0.10$), suggesting that the numerical setup reproduces a statistically more homogeneous atmosphere than the real quiet Sun, lacking the full magnetic and radiative complexity that diversifies the photosphere–chromosphere relationship in observations. This is confirmed by the substantially higher JSD values obtained for the granular regions (D and E) in the observational case (panel b), where JSD reaches 0.41-0.42. 
 
 The JSD of 0.17 at the simulation footpoint remains relatively low overall and is consistent with a coherent vertical transport channel in which reflection, dissipation, and mode conversion progressively modify the signal with height while still preserving a significant degree of shared structure between atmospheric layers. In particular, this vortex has previously been shown to enhance dissipation \citep{Silva_2024a}, likely contributing to the moderate divergence between the photospheric and chromospheric perturbation distributions.  

In the simulation, the enhanced statistical coupling of Region A, indicated by the highest NMI among all sampled regions, is supported by the known physical connection via velocity and magnetic field lines linking the photospheric vortex footpoint to the chromospheric swirl, together forming a coherent three-dimensional vortical structure \citep{Silva_2024a}. This provides a physically motivated and simulation-supported criterion for footpoint identification in observational data: since full three-dimensional velocity fields across the solar atmosphere remain observationally inaccessible, the vortex footpoint can be identified as the photospheric region exhibiting the highest NMI with the overlying chromospheric swirl. JSD serves as a complementary diagnostic of the physical nature of the coupling, which helps characterize how the signal is modified during propagation across the solar atmosphere.

In the observations, Region A exhibits substantially greater shared information than all other sampled regions, with NMI increases of approximately 17\%, 50\%, 85\%, and 183\% relative to regions B (bright point), D (granular region), E (granular region), and C (bright point), respectively. By analogy with the simulation, where the highest NMI uniquely identified the magnetically connected footpoint, the comparably elevated NMI of Region A suggests that chromospheric swirl S1 is magnetically rooted in Region A over the analysed time interval, with vortex-driven perturbations propagating coherently along a shared flux tube.

To assess whether the enhanced NMI at the candidate footpoint reflects a preferred photospheric-chromospheric temporal association, we performed a circular-shift surrogate test. For each candidate region, the chromospheric time series was shifted relative to the photospheric series over all non-zero temporal offsets, preserving the temporal structure of each signal while changing their relative alignment. To ensure consistency with the time interval used in the SPOD analysis, the simulation interval was restricted to $T=23$ frames, yielding $T-1=22$ distinct non-zero circular shifts, all of which were evaluated. We define
\begin{equation}
    p_{\rm emp} = \frac{r+1}{N+1},
\end{equation}
where $r$ is the number of shifted realizations with NMI equal to or greater than the observed value and $N$ is the total number of shifts tested. For the simulation, where the physical connectivity is independently known, Region A has both the highest NMI and the largest positive departure from its surrogate distribution ($z=1.02$, $p_{\rm emp}=3/23\approx0.130$).By contrast, Region D, which has the second-highest NMI in the simulation but no physical connectivity to the chromospheric swirl, shows no positive departure from its surrogate distribution ($z=-0.74$, $p_{\rm emp}=18/23\approx0.783$), with 17 of the 22 shifted realizations reaching or exceeding the observed NMI.
 Although Region A does not reach conventional statistical significance, it shows the strongest positive departure from the temporal-shift null compared with Regions B-E in the simulation, consistent with its independently known physical connectivity. The remaining regions show weaker or no such departure ($z=-0.74$-$0.82$, $p_{\rm emp}=0.227$-$0.773$). 

The same test was applied to the observations, for which $T=21$ frames provide 20 distinct non-zero shifts, Region A shows both the highest NMI and the largest positive departure from its surrogate distribution ($z=1.48$, $p_{\rm emp}=1/21\approx0.048$): none of the 20 shifted realizations produced an NMI as large as that obtained for the observed alignment. Although Region B has the next-highest NMI, it is not exceptional relative to its temporal-shift null distribution ($z=0.43$, $p_{\rm emp}=0.40$), indicating that its apparent coupling is consistent with that produced by temporal misalignment. Region C is also associated with a photospheric bright point but has a considerably lower NMI and shows no evidence of enhanced coupling relative to its null distribution ($z=-1.58$, $p_{\rm emp}=1.00$).

Beyond the NMI criterion, Region A exhibits the lowest JSD in the observational case (swirl S1), indicating that its photospheric fluctuations are statistically most similar to those of the chromospheric swirl. The near-zero JSD at the observed footpoint suggests that, unlike in the simulation, perturbations generated at the vortex footpoint undergo negligible statistical transformation during propagation along the flux tube, consistent with a more linear or weakly dissipative coupling regime in the observed vortex compared to its simulated counterpart.

The other bright point in the observational case, Region C, shows a comparable JSD to Region A, likely because both bright points lie along the same intergranular lane, where they are subject to similar perturbations driven by the same granular convective pattern. However, its substantially lower NMI indicates that Region C's perturbations are not coherently coupled to the swirl's dynamics, distinguishing a shared background environment from a direct physical connection to the vortex.

Region B, also a bright point, presents a contrasting case: despite relatively high NMI (48.5\%), its shared information is approximately 14.6\% lower than that of Region A (56.8\%), and its substantially elevated JSD (0.36) suggests a more indirect connection to swirl S1. The photospheric fluctuations in Region B convey information about the swirl dynamics, but with substantially modified amplitude statistics. This is consistent with coupling through the swirl's surrounding magnetic canopy rather than a direct flux-tube connection to its core

The control regions D and E, sampling granular structures, exhibit both reduced NMI and elevated JSD. The reduced NMI is consistent with weaker magnetic connectivity to the swirl, while the elevated JSD likely arises from local convective dynamics that decorrelate photospheric and chromospheric signals at these locations.

\subsection{MHD wave analysis}

\subsubsection{Wave mode detection}
After establishing photosphere-chromosphere vortex coupling, we applied SPOD to observational, numerical and synthetic data to isolate and characterise coherent wave modes across atmospheric heights.
Its classical version, POD, has been successfully applied to solar magnetic structures, such as sunspots, to identify multiple high-order eigenmodes in the photosphere \citep{Albidah2021, Albidah_2023, Jafarzadeh_2024}.

Mode selection was based on the identification of modes with time coefficients that presented periodic behaviour and spatial modes with patterns that presented kink or sausage mode signatures (see Appendix \ref{app:spod-model} for more details). We do not search for torsional Alfven waves as perturbations of intensity line-core traces fluctuations in chromospheric mass density \citep{Leenarts_2012} and thus are inherently sensitive to compressible perturbations only, naturally excluding such waves.

\begin{figure*}
    \begin{center}
    \includegraphics[width=1\textwidth]{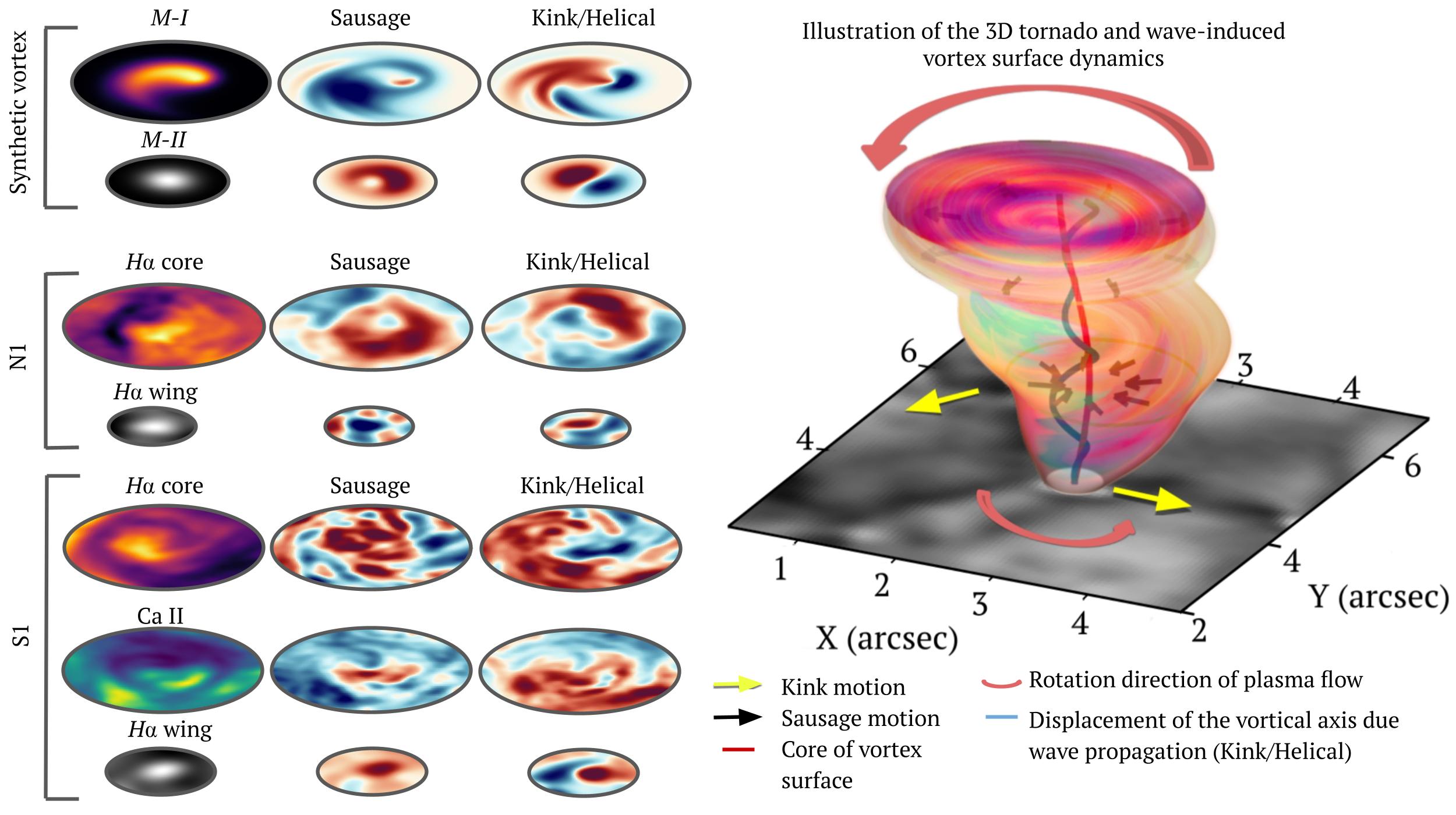}
    \end{center}
    \caption {Detected wave modes. Left panel: Spatial distribution of dominant wave modes in a solar tornado structure, categorised by data type: modelled vortex (M-I and M-II), numerical simulation (vortex N1), and observational data (vortex S1). The intensity maps of the vortex are shown in the first column, while each subsequent column corresponds to a different type of MHD wave: sausage modes, and kink/helical modes. For clarity, the photospheric counterpart of the vortex is shown at twice its actual size, and the vortex cross-sections are displayed in alignment. The radii of vortex regions for S1 is 0.22 arcseconds (photosphere - H$\alpha$ wing) and 0.71 arcseconds (chromosphere - Ca II and H$\alpha$ core). For vortex N1, the radii are 0.26 (photosphere - H$\alpha$ wing)	and 1.2 arcseconds (chromosphere - H$\alpha$ core)
 Right panel: Three-dimensional schematic representation of a tornado-like vortex illustrating the co-existence of sausage and helical wave modes propagating along a rotating plasma flux tube. The model is based on measured parameters from S1, rooted in the photosphere (H$\alpha$ wing intensity) and extending into the chromosphere (H$ \alpha$ core). The coloured surface shows the oscillatory displacement amplitude derived from SPOD.
}
\label{fig:detectedmodes}
\end{figure*}

Figure~\ref{fig:detectedmodes} displays the spatial structure of MHD wave modes identified through SPOD using a combination of vortex-model data, numerical simulation, and high-resolution observations from SST/CRISP of solar tornadoes. The vortex-model data show that the appearance of wave modes is strongly influenced by plasma rotation, particularly under differential rotation (M-I). Comparison with H$\alpha$ core and wing observations of vortices S1 and N1 indicates that chromospheric swirl patterns closely resemble M-I, whereas photospheric bright-point structures are more consistent with solid-body rotation (M-II).

In the vortex-model dataset, the physical components are explicitly prescribed, consisting of background rotation, superimposed kink and sausage waves, and low noise. This controlled setup enables a direct interpretation of the SPOD results. The decomposition isolates coherent structures from noise, which appears mainly in higher-order modes. Thus, the dominant modes primarily reflect the imposed dynamics, namely the wave motions and the rotational background.
The resulting mode structure deviates from the classical expectations for cylindrical waveguides. Such differences are anticipated, given that traditional models typically neglect critical physical aspects included in our setup, such as background rotation and elliptical cross-section geometry \citep{Aldhafeeri_2021, Skirvin_2023}. 
For observational cases, the waveguide and realistic plasma density distributions can change the appearance of wave signatures in SPOD modes even further \citep{Ballai_2024}. 
Our analysis of M-I and M-II demonstrates that SPOD reliably recovers all imposed wave modes despite added noise and random motion.

As SPOD assumes linear superposition of modes, we applied a Gaussian low-pass spatial filter to the N1 H$\alpha$ data, smoothing structures smaller than 0.3\arcsec\ to reduce small-scale nonlinear signatures and maintain consistency with the method's assumptions. For the SPOD analysis, we used a time interval in which the vortex was well established, displaying clear rotational features in both the photosphere and the chromosphere. For S1, our analysis was from 10:21\,UT to 10:24\,UT between photosphere and Ca\ {\sc ii}\ and 10:21\,UT to 10:28\,UT to H$\alpha$ core. Since bright points show considerable movement, 
SPOD was restricted to time windows short compared to the total lifetime of each structure. Because SPOD is variance-optimal rather than physics-aware, it is intrinsically sensitive to advection and non-stationarity, motivating the use of short analysis windows in the presence of significant feature motion.

Any time window selected for the photosphere will display both the sausage and kink modes. For vortex N1, the analyzed interval spans 500 s from $t = 3800$ s. Although SPOD reveals additional structures potentially associated with more complex wave behavior in cylindrical or elliptical geometries, our analysis focuses on sausage, kink, and helical modes, which are most robust to density inhomogeneity and waveguide irregularities \citep{Aldhafeeri_2021,Ballai_2024}. The presence of these modes across different atmospheric heights indicates that the tornado-like structure acts as a multi-layered MHD waveguide of magnetoacoustic waves in the lower atmosphere. 

Notably, all identified sausage modes correspond to overtone structures.
For the S1 modes, it is evident that the kink mode present in the photosphere evolves into a slightly twisted kink, i.e. a helical mode, at the Ca\ {\sc ii} height and ultimately develops into a full helical mode in the H$\alpha$ core; consistent with kink-rotation coupling that reproduces the characteristic helical wave modes known from hydrodynamic vortex theory \citep{Maxworthy_Hopfinger_Redekopp_1985}. Thus, for the rest of the paper, we refer to those modes as kink/helical as they present themselves as a kink in the photosphere.

The right-hand side of the figure presents a three-dimensional schematic picture for the case in which sausage and helical modes propagate along a rotating plasma column, which is anchored in the photosphere and extends into the chromosphere. Parameters such as frequency, displacement, and expansion/contraction due to waves were obtained by the analysis presented in 
Fig.~\ref{fig:simmodes} in the Appendix. The coloured surface shows how SPOD-identified waves shape the vortex: sausage modes modulate the cross-sectional area of the vortex tube, while rotating kink modes impose a lateral displacement that curves the structure, spatially linking the photospheric bright point to the chromospheric swirl. The vortex thus emerges as a three-dimensional coherent MHD waveguide of magnetoacoustic waves whose wave-driven bending enables inter-layer coupling and produces its tornado-like morphology in the lower atmosphere.

\subsection{Wave identification from morphological analysis}
To cross-validate the wave detection by SPOD, we conducted an independent analysis based on the position of the vortex center and its shape over its lifetime; i.e, a morphological analysis. The center of the vortex and its boundary were identified using the A-MorphIS code \citep{Dakanalis_2021, Dakanalis_2022}. For the observational data, A-MorphIS was applied to CRISP Ca\ {\sc ii} line center and H$\alpha$-0.2~\AA\ intensity filtergrams, while for simulation data, temperature gradients were used as input. The temperature gradient was chosen because it provided a good proxy for the vortex boundary as shown by \cite{Silva_2024a}. 

\begin{figure*}
         \includegraphics[width=\textwidth]{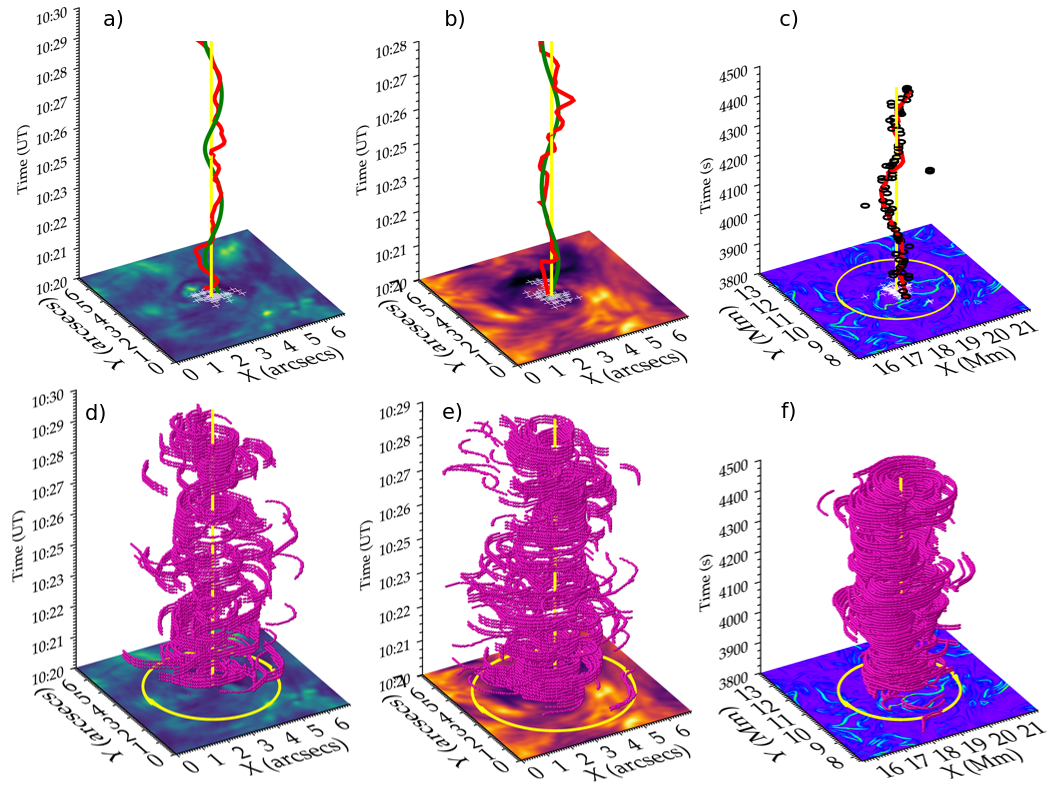}
     \caption{Detection of helical (top row) and sausage (bottom row) modes in vortex region of observational (a, d for Ca\,{\sc II} and b, e for H$\alpha$) and simulation vortex (c, f). The vertical axis represents the Bifrost and CRISP timescales, starting at $t=3800$ s and $t=10$ : 20 UT, respectively. The base  image corresponds to the initial H$\alpha$ emission of S1 and to temperature gradient in the field of view of N1.  Purple lines show the derived vortex segments with the A-MorphIS code. The yellow circles and vertical lines represent the swirl radii, calculated from the outermost 10\% segments, and the detected mean centers, respectively. The red line represents the curve fit of the center positions (black small circles), and the green line is a helical curve fitted to the positions of the center.}
 \label{fig:simmodes}
\end{figure*}

The temporal evolution of the studied swirls is presented in Fig.~\ref{fig:simmodes} with CRISP observations in the top row and Bifrost simulation results in the bottom row. These findings cross-validate and further support the proposed wave propagation modes. From the morphological analysis, we computed that the frequencies for both modes increase with height, from 2.88$\pm$0.38 to 3.82$\pm$0.97 mHz for kink/helical-like modes, and from 4.38$\pm$1.55 to 4.98$\pm$1.3 mHz for sausage-like modes (see Table~\ref{tab:freq} in the Appendix \ref{app:morphologic-wave} for other vortices), and are consistent with previously reported wave frequency ranges within vortices  \citep{Tziotziou_2019}. The mode frequencies differ between layers, consistent with the expected evolution of waves propagating through a stratified atmosphere.

\begin{figure*}[htb!]
\centering
{\includegraphics[width=0.45\textwidth]{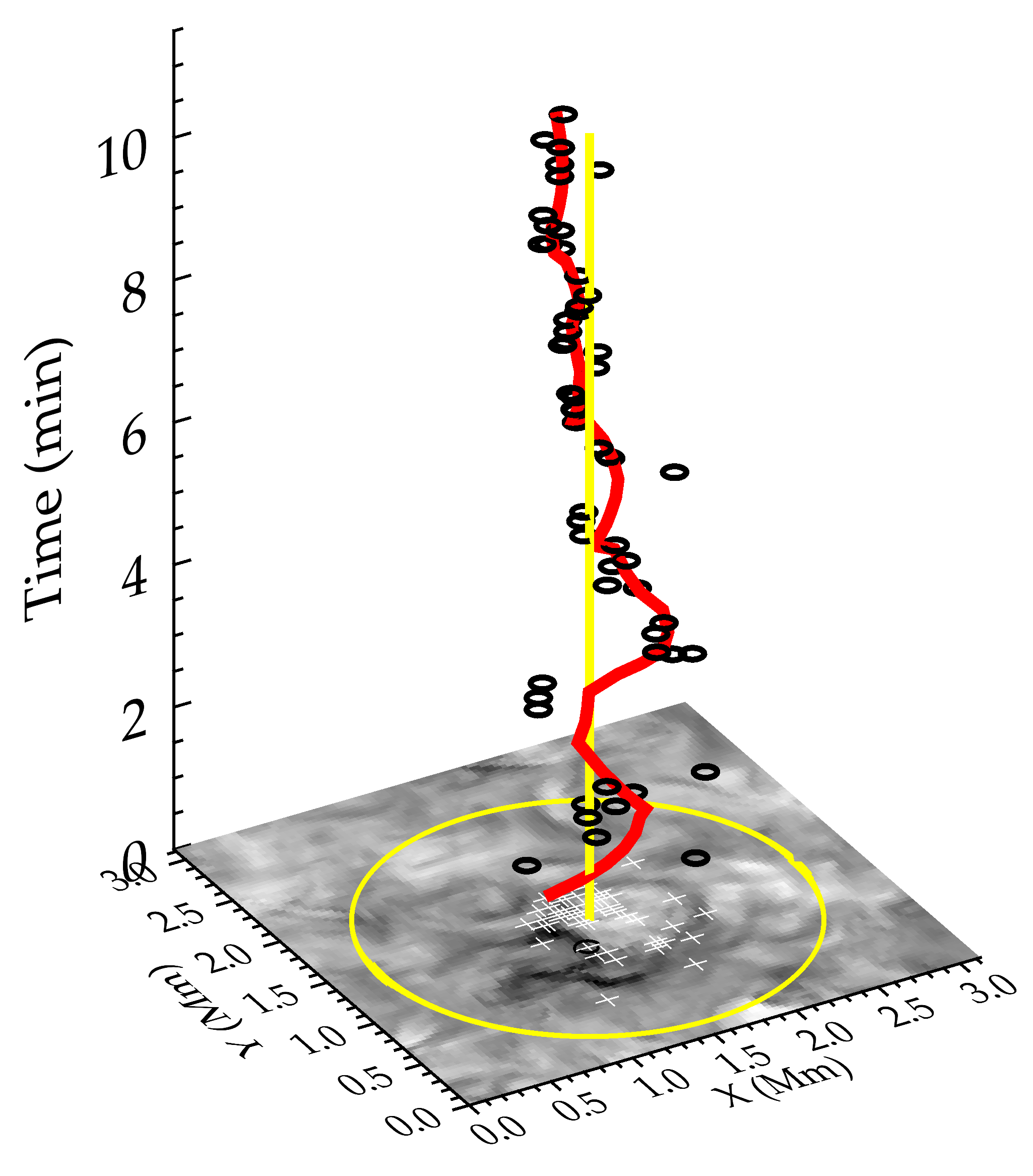}}
{\includegraphics[width=0.45\textwidth]{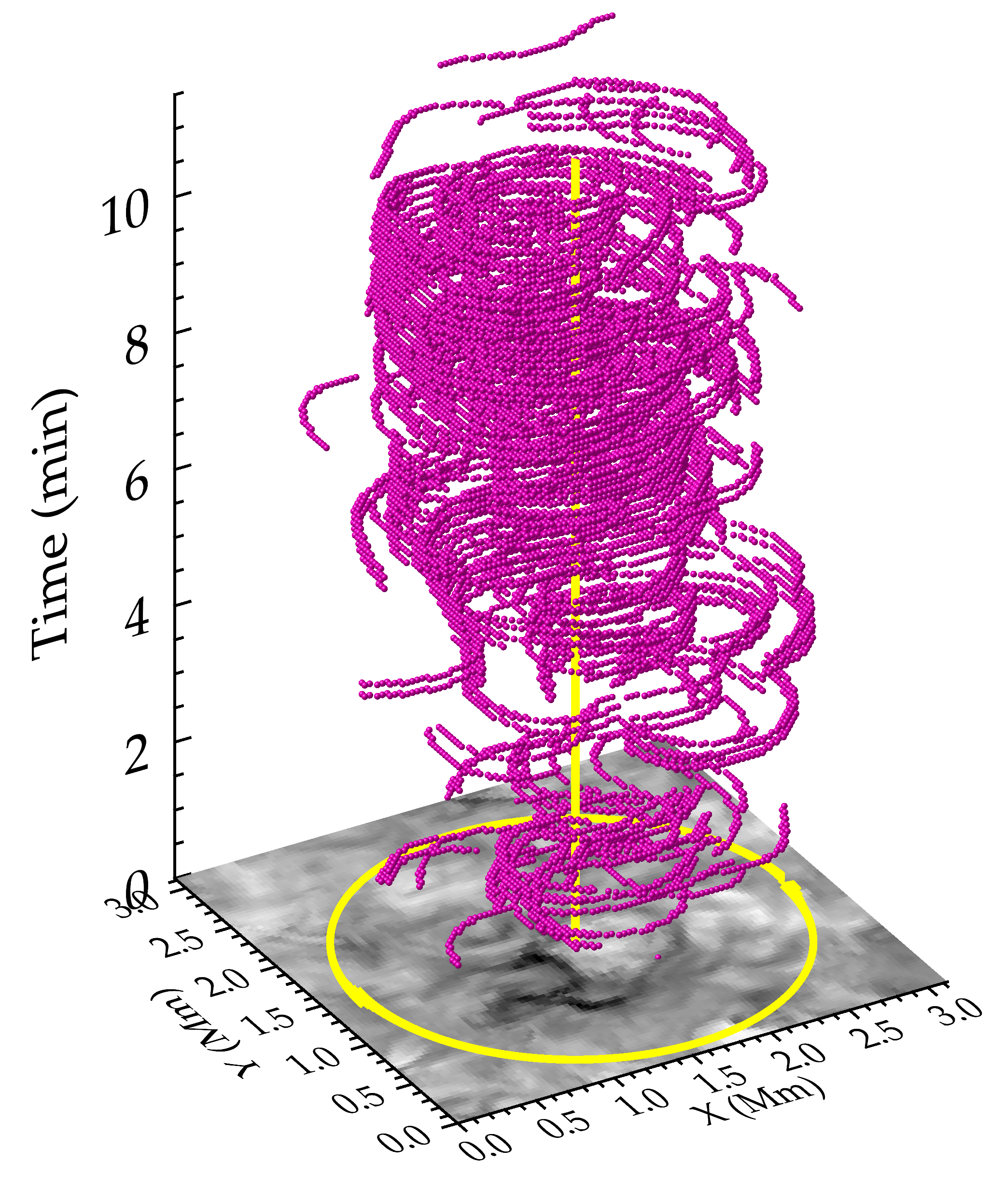}}
\caption{Detection of helical (left panel) and sausage (right panel) modes in the synthetic H$\alpha$ intensity of the simulation vortex at the approximated formation height of the H$\alpha$ line center. The vertical axis represents the Bifrost timescale, starting at $t=3800$ s. The base grey-scale image corresponds to the initial H$\alpha$ emission in the field of view. Purple lines show the derived vortex segments using the A-MorphIS code, while yellow circles and vertical lines represent the swirl radii, calculated from the outermost 10\% segments, and the detected mean centers, respectively. The red line represents the curve fit of the center positions (black small circles). 
}
\label{fig:sw}
\end{figure*}

A similar analysis to that shown in  
Fig.~\ref{fig:simmodes} has been performed on the simulated vortex, focusing on the evolution of its center's position and shape over time. However, the A-MorphIS code has now been applied to synthetic H$\alpha$ intensities at the approximate formation height of the line center. The corresponding results are displayed in Fig.~\ref{fig:sw}. 
The visual differences between Figs.~\ref{fig:simmodes} and \ref{fig:sw} are mainly due to the simulated vortex being less distinct in synthetic H$\alpha$ intensity than in temperature gradients, which seem to more effectively capture the underlying vortex dynamics. 
Nevertheless, the intensity-based results support and cross-validate the wave propagation modes previously identified through the temperature gradient analysis. A detailed morphological analysis for the other observational swirls is presented in Appendix \ref{app:morphologic-wave}.

\subsubsection{Wave-driven dynamics}
\begin{figure*}
\centering
    \includegraphics[width=0.85\textwidth]{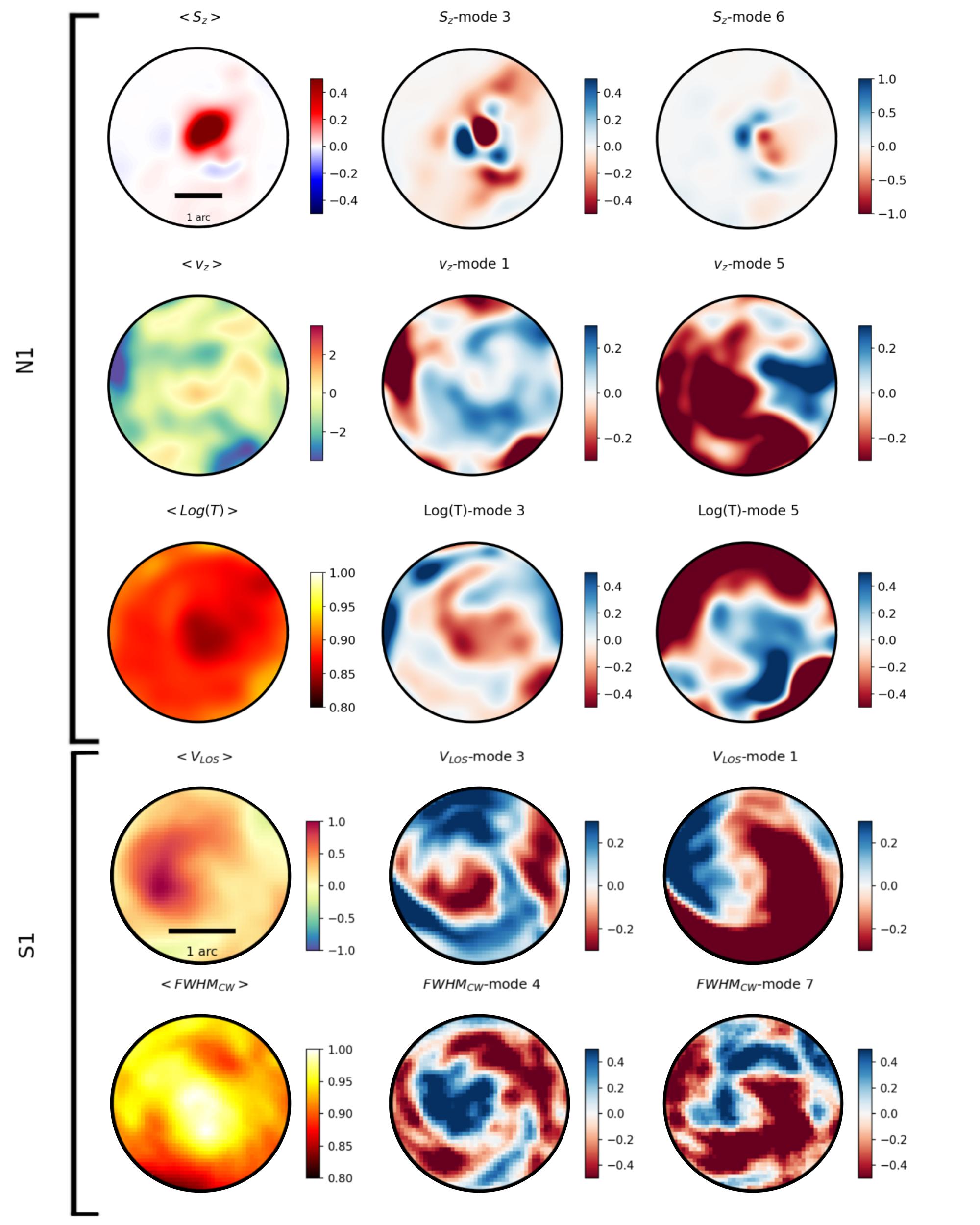}
    \caption{SPOD analysis of plasma variables and observables. The first column shows the variable average for the time interval of the analysis ($t=3800$ s to $t=4300$ s for N1 and {from 10:20UT to 10:27UT} for S1). For the vortex N1, we compute the SPOD modes of $S_z$,  $v_z$ and $Log(T)$, whereas for S1, we analyze $v_{LOS}$ and the FWHM core width, which works as a proxy for temperature. The last two columns depict the space modes of SPOD that were identified as MHD wave modes: sausage (second column) and kink/helical mode (third column). The SPOD mode amplitudes in second and third column were normalised to the interval $[-1,1]$ and saturated to facilitate mode identification.
}
\label{fig:comparemodes-main}
\end{figure*}

To assess how the detected wave modes influence overall plasma dynamics, we extended the SPOD analysis to include key physical quantities with the results presented in Fig.~\ref{fig:comparemodes-main}. For the numerical simulation (vortex N1), we examined the vertical Poynting flux ($S_z$), vertical velocity ($v_z$), and temperature, while for the observational case (vortex S1), we analysed the line-of-sight velocity ($v_{\mathrm{LOS}}$) and the full width at half maximum (FWHM) of the H$\alpha$ core, the latter serving as a proxy for chromospheric temperature \citep{Leenarts_2012}. For S1, all analyses were performed over the same time interval used in 
Fig.~\ref{fig:detectedmodes}. However, for N1, the temperature only shows the wave modes with a given delay, so we shift our analysis by 100 seconds forward.

The first column in Fig.~\ref{fig:comparemodes-main} shows time-averaged, height-integrated maps of the relevant variables. For N1, values are extracted at the H$\alpha$ core formation height, enabling assessment of wave-driven effects at the line-formation layer. The last two columns display the dominant SPOD spatial modes over the same interval; the colour scale is saturated to facilitate mode identification.
For N1, the SPOD modes of $S_z$, $v_z$, and temperature primarily display kink and sausage overtone structures. These wave signatures in the vertical Poynting flux indicate that compressible MHD waves strongly mediate energy transport. 

In the observational case (S1), SPOD analysis reveals a stronger presence of helical modes in both $v_{\mathrm{LOS}}$ and FWHM. The $v_{\mathrm{LOS}}$ SPOD modes show spatial patterns consistent with the dynamics of helical and sausage waves, suggesting that these waves contribute to  vertically coherent plasma motions. Specifically, the dominance of helical and sausage modes suggests that vertical energy transfer occurs via structured, compressible dynamics. 

\subsubsection{Wave energy generation and propagation}
The wave energy flux can be written as \citep{Mumford_2015}
\begin{equation}
\mathbf{W}_{wave} = p_k' \mathbf{v} + \frac{1}{\mu_0} (\mathbf{B}_b \cdot \mathbf{B}') \mathbf{v} - \frac{1}{\mu_0} (\mathbf{v} \cdot \mathbf{B}') \mathbf{B}_b,
\end{equation}
where \( \mathbf{v}\) is the velocity, \( p_k' \) is the kinetic pressure perturbation, \( \mathbf{B}_b \) is the background magnetic field, and \( \mathbf{B}' = \mathbf{B} - \mathbf{B}_b \) is its perturbation. This expression is split into compressive (pressure, $p$) and magnetic ($m$) components by writing
\begin{align}
\mathbf{W}p &= p_k' \mathbf{v}, \\
\mathbf{W}m &= \frac{1}{\mu_0} (\mathbf{B}_b \cdot \mathbf{B}') \mathbf{v} - \frac{1}{\mu_0} (\mathbf{v} \cdot \mathbf{B}') \mathbf{B}_b.
\end{align}
The compressive ($\mathbf{W}p$) and magnetic ($\mathbf{W}m$) components of the wave energy flux are computed for the 10 vortices  shown in Fig.~\ref{fig:useddomains}.  {To assess the specific impact of vortex dynamics on wave energy flux, we compared the average flux in vortex regions with that in equally sized, non-vortex regions (see Fig.~\ref{fig:ratios}). 
Across the atmosphere, vortical structures amplify magnetic wave energy transport by roughly a factor of three, consistent with a waveguiding effect. The compressive component exhibits a weaker but still significant enhancement in the upper atmosphere, reaching a factor of two near $\sim$2 Mm in the low corona.}

{The vertical magnetic-to-pressure flux ratio $Wm_{z}/Wp_{z}$, as Fig.~\ref{fig:ratios} shows, increases monotonically with height. Within vortical structures, the ratio $Wm_{z}/Wp_{z}$ spans over four orders of magnitude, increasing from below $10^{-3}$ near the photosphere to above $10$ at heights of $\sim$2 Mm. This indicates that, in vortices, the compressive wave energy dominates at low altitudes, while magnetic energy becomes increasingly more important higher in the atmosphere. Further analysis on wave energy transport is presented in Appendix \ref{app:waveflux}.

The kink/helical and sausage modes can perturb both density and magnetic field; therefore, they can contribute to both $Wp_z$ and $Wm_z$. Consequently, the identification of a helical mode at the H$\alpha$ core does not imply that the energy transport is already magnetically dominated. The result $Wm_z/Wp_z \sim 1$ at this height simply indicates that compressive and magnetic contributions are of comparable importance. The subsequent increase to $Wm_z/Wp_z > 1$ at greater heights suggests that magnetic energy transport becomes progressively more important, but does not require a change in the identified modes.

At $2\,\mathrm{Mm}$ above the solar surface, vortex N1, with a radius of
$\approx1.5''$ ($\approx1094\,\mathrm{km}$) and a lifetime of
$\sim500\,\mathrm{s}$, transports a total vertical wave energy of
$\sim1.3\times10^{25}\,\mathrm{erg}$ through the upper atmosphere
(see Appendix \ref{app:waveflux}, Fig.~\ref{fig:wave_fluxes}). This
energy is estimated from the mean vertical wave-energy flux,
$W_{{\rm wave},z}=W_{p,z}+W_{m,z}$, over the horizontal area of the
vortex and its lifetime. Of this,
$E_{m,z}\simeq9.4\times10^{24}\,\mathrm{erg}$ is associated with the
magnetic component and
$E_{p,z}\simeq3.8\times10^{24}\,\mathrm{erg}$ with the compressive
component. Thus, despite the increasing predominance of the magnetic
component above 1 Mm, the compressive component still contributes a
substantial fraction of the wave energy transported vertically through
the upper atmosphere. The total energy transported by N1 remains of
order $10^{25}\,\mathrm{erg}$, placing the upward wave-energy transport
associated with an individual vortex within nanoflare-like energy scales.

In the Bifrost simulation, the lower boundary condition is consistent with that used in the enhanced network 
simulation \citep{Carlsson2016}, where the lower boundary is a pressure node that reflects acoustic waves. The issue is that there is a limited set of excited p-modes, which are then assigned much larger amplitudes than in the real Sun. This could artificially enhance compressive wave energy flux ($Wp$), especially for global-scale modes, and complicates direct interpretation of $Wp \gg Wm$ below 1~Mm as a purely physical result. While the lower boundary condition may contribute to elevated $Wp$ in the lower atmosphere, the correlation between compressive wave activity and temperature variability in both simulations and observations strongly suggests that compressive modes naturally play an important role in chromospheric energy transport and heating.

\begin{figure}
    \centering
    \includegraphics[width=\columnwidth]{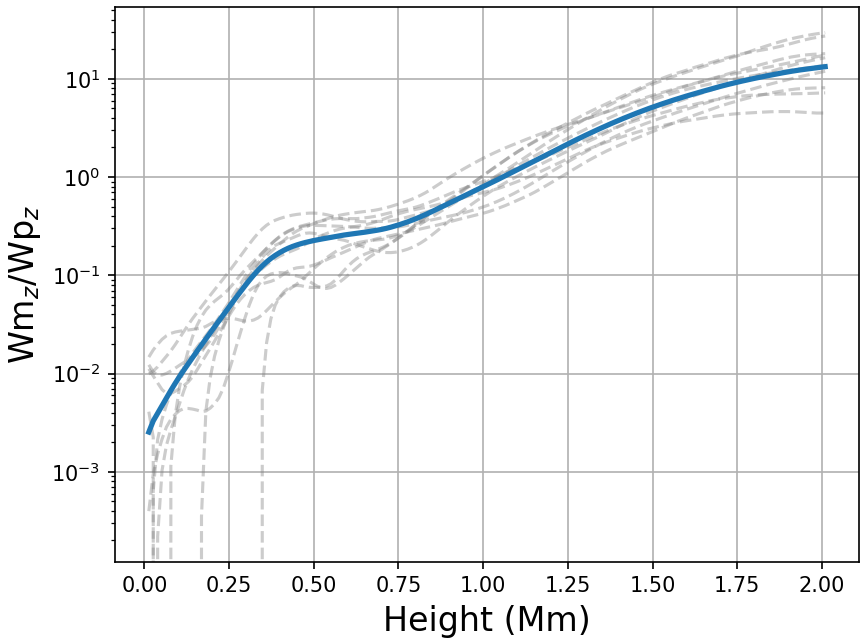}
    \includegraphics[width=\columnwidth]{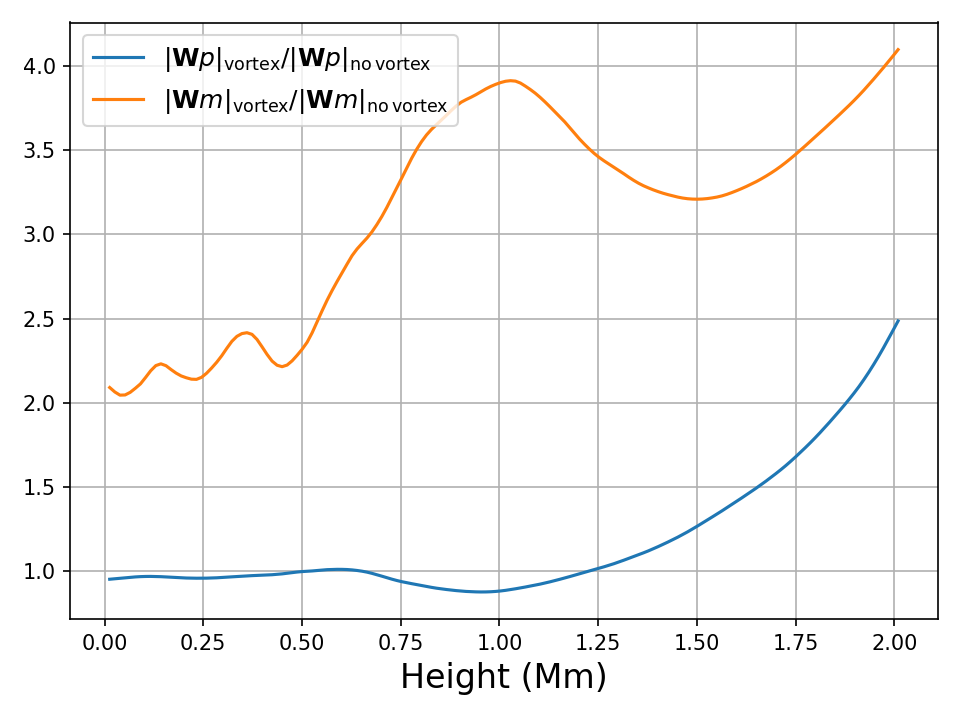}
    \caption{
    Top panel: Height dependence of the ratio between the vertical magnetic and pressure wave-energy fluxes, $W_{m,z}/W_{p,z}$, within the vortex regions. Grey dashed lines show the ratio for each individual vortex, while the blue line shows the mean of the individual-vortex ratios, $\langle W_{m,z}/W_{p,z}\rangle$. Bottom panel: Mean energy-flux ratio in vortical and non-vortical regions as a function of height.}
    \label{fig:ratios}
\end{figure}

\section{Discussion and Conclusions}

Combining high-resolution observations and realistic MHD simulations, we investigate the role of solar vortices as waveguides in the lower atmosphere. Grounded in the known three-dimensional magnetic connectivity of the vortical structure from numerical simulations, we provide a quantitative assessment of vertical coupling using NMI and JSD. Applying these diagnostics to observational solar vortices enabled the identification of the photospheric footpoint associated with the chromospheric swirl.  

NMI quantifies the shared information between atmospheric layers, while JSD measures the statistical similarity of their fluctuation distributions. Together, they provide complementary diagnostics of photosphere–chromosphere coupling: NMI captures the existence and strength of the connection, while JSD characterises how the signal is reshaped during propagation. In the simulation, the photospheric vortex footpoint (independently determined from three-dimensional vortex detection \citep{Silva_2024a}) shares 47.8\% of its information with the chromospheric swirl, representing an 11.3-76.9\% relative enhancement over control regions. In the SST observations, Region A attains 56.8\% shared information with the chromospheric swirl, exceeding all other sampled regions by 17-183\%. The JSD at the simulation footpoint (0.17) indicates that perturbation statistics are moderately reshaped during propagation. This is consistent with a coherent vertical channel in which reflection, dissipation, and mode conversion progressively modify the signal with height. By contrast, the near-zero JSD at the observed footpoint (0.05) suggests that perturbations are transmitted with negligible statistical transformation, indicative of a more linear or weakly dissipative coupling regime. 

The presence of similar dominant SPOD modes at photospheric and chromospheric heights suggests that the tornado-like structure behaves as a multi-layered wave-supporting structure in both simulations and observations. We therefore use the term MHD waveguide in a phenomenological sense: the coherent perturbations remain spatially associated with the magnetised vortex region across multiple heights. Similar kink/helical and sausage-like disturbances have been reported in hydrodynamic vortex flows in the absence of magnetic fields \citep{dahl2021centrifugal}, indicating that vortical flows can support and organise such perturbations. At the same time, solar vortices are embedded in a magnetised, stratified plasma, and the magnetic field associated with vortical structures need not remain confined to the same vortex region at all atmospheric heights \citep{yadav_2026}. The effective guiding structure should therefore be understood as the combined action of vortical flow, magnetic-field organisation, thermodynamic structuring, and transverse gradients, rather than as either an isolated flow vortex or an isolated magnetic flux tube.

{ Our results, obtained for both simulation and observational data, demonstrate that magnetic vortices naturally support coherent magnetoacoustic wave modes, highlighting their role as robust, self-organising channels for wave energy transport in the lower solar atmosphere. The dominant SPOD modes show consistent signatures of MHD waves in cylindrical and elliptical waveguides. This contrasts with what would be expected from irregular or unstructured oscillations. Processes such as turbulent displacements or random, non-modal responses typically do not project strongly onto the leading spatial modes. Instead, they tend to appear in higher-order modes, where they manifest as broadband, spatially incoherent signatures. Instead, we observe distinct, large-scale structures consistent with propagating wave behaviour.}

In the lower solar atmosphere, our analysis identifies signatures of kink/helical and sausage modes in the vortex region, consistent with compressive magnetoacoustic wave activity. The transition from $Wm_{z}/Wp_{z}<1$ to $Wm_{z}/Wp_{z}>1$ marks a change from compressive-energy-dominated to magnetic-energy-dominated transport, occurring above the H$\alpha$ formation height ($\approx 1$~Mm in this simulation). This might indicate mode conversion, but it cannot exclude the alternative possibility that compressive waves shock and dissipate faster than the magnetic perturbations, consistent with chromospheric heating \citep{yadav_2026}.

Several earlier studies have identified torsional Alfv\'en waves in chromospheric vortices based primarily on the morphological appearance of the magnetic field and on propagation speeds. \citet{Wedemeyer2012} interpreted co-rotating magnetic field lines as torsional Alfv\'en waves due to their visual appearance and apparent upward propagation at the Alfv\'en speed. Similarly, \citet{Liu_2019} and \citet{Battaglia_2021} reported Alfv\'enic pulses in vortex structures below 1 Mm based on twisted magnetic-field configurations and pulse propagation speeds consistent with the Alfv\'en velocity. None of these studies, however, separated the magnetic perturbation from the total magnetic field or examined whether the transverse perturbations satisfy the Wal\'en relation,
\begin{equation}
 \mathbf{u}^\prime = \pm \frac{ \mathbf{B}^\prime}{\sqrt{\mu_0 \rho}},
\label{eq:walen}
\end{equation}
where the sign depends on the propagation direction relative to the background magnetic field \citep{cable_99}.

We therefore analysed the perturbation components perpendicular to the local background magnetic field. To extract these components, we performed the SPOD decomposition using a three-dimensional vectorial formulation in which the covariance matrix is constructed from the magnetic-energy norm. This is a physically motivated choice because Alfv\'en waves involve magnetic-field perturbations, and a basis that maximises coherent magnetic-energy structures should be sensitive to such signatures.

Figure~\ref{fig:bmodes} shows the SPOD mode 1 perturbation components perpendicular to the local background magnetic field at selected heights. Up to 1~Mm, these perturbations do not exhibit a clear torsional pattern, and the Wal\'en ratio departs from the Alfv\'enic expectation of unity. Near the photosphere, at $z\approx0.2$~Mm, it reaches values of $\sim5$,indicating that the velocity perturbation is substantially larger than expected from the accompanying magnetic perturbation for an Alfv\'en wave.

At $0.6$ and $1$~Mm, the ratio remains below unity and varies markedly throughout the vortex evolution, with mean values of $0.59$ and $0.29$, respectively. Above approximately $1$~Mm, a torsional Alfv\'en-like pattern emerges. However, the Wal\'en ratio only intermittently reaches values of approximately 0.7  as shown in the corresponding animation of the figure, but remains generally below unity as the provided average values indicate}. Thus, although the transverse perturbations shown by the arrows are compatible with a torsional wave-like disturbance above 1~Mm, the velocity and magnetic-field perturbations are not persistently balanced as expected for a pure Alfv\'en wave.   In other words, the Wal\'en-ratio behaviour does not support a definitive Alfv\'enic perturbation/wave identification across the analysed heights. The departure from the ideal Wal\'en relation may reflect nonlinear vortex dynamics, height-dependent background structuring, or mixed-mode behaviour, and should be investigated in future studies.
\begin{figure*}
\centering
\includegraphics[width=1\linewidth]{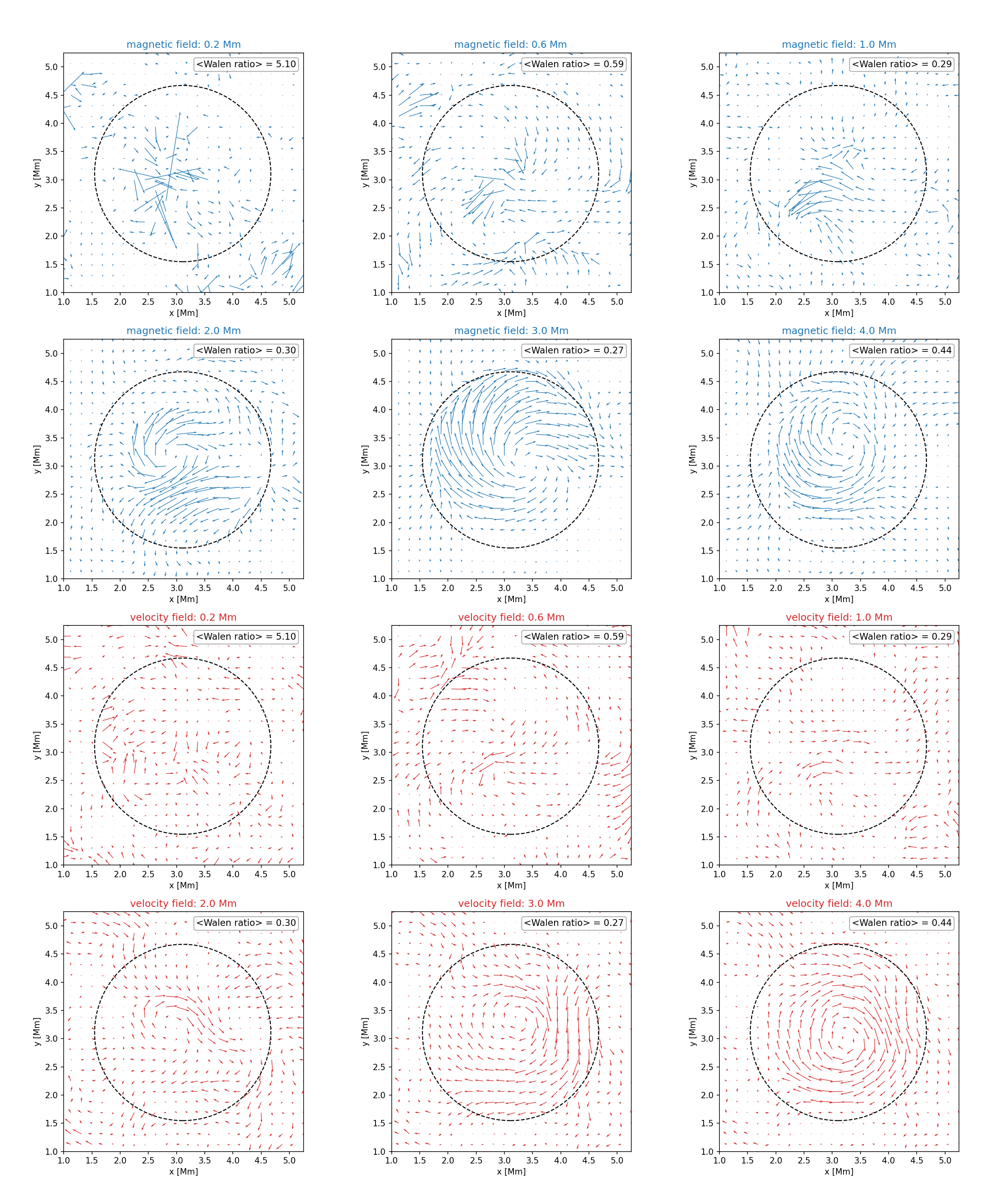}
\caption{
Vectorial SPOD analysis of the N1 vortex during the wave-analysis interval. 
Each panel shows the in-plane perturbation vectors of the reconstructed magnetic-energy SPOD mode, projected onto the plane perpendicular to the local background magnetic field, at six heights: 0.2, 0.6, 1.0, 2.0, 3.0, and 4.0~Mm. 
The top two rows (blue) show the magnetic-field perturbation, $\delta\mathbf{B}_{\perp}$, while the bottom two rows (red) show the velocity-field perturbation, $\delta\mathbf{u}_{\perp}$. 
The dashed circle marks the vortex core over which the Wal\'en amplitude ratio,
$\left\langle |\delta\mathbf{u}_{\perp}| / |\delta\mathbf{B}_{\perp}/\sqrt{\mu_0\rho}| \right\rangle$,
is computed; its time-averaged value is annotated in the top-right of each panel. An animation of this figure is \href{https://drive.google.com/file/d/10BvSEE5USpqn8gqFygU5FgL64natmKK_/view?usp=sharing}{available here}.}

\label{fig:bmodes}
\end{figure*}

The apparent discrepancy between our findings and those of \citet{Liu_2019} and \citet{Battaglia_2021}, who reported Alfv\'enic pulses in vortex structures below approximately 1 Mm, likely reflects differences in the criteria used for wave identification. In this height range, our analysis does not find clear signatures consistent with a torsional Alfv\'en wave. These earlier studies identified Alfv\'enic pulses primarily from twisted magnetic-field configurations and propagation speeds comparable to the Alfv\'en velocity. In contrast, we isolate the magnetic perturbations perpendicular to the local background field using SPOD mode decomposition and further evaluate their consistency with the Wal\'en relation. Over the same height range, we find neither a clear torsional perturbation pattern nor a Wal\'en ratio close to unity: the ratio is strongly elevated near the photosphere ($\sim5$ at $z\approx0.2$~Mm) and remains  below unity at other heights. Under these more restrictive criteria, the perturbations up to approximately 1~Mm are therefore not consistent with an unambiguous torsional Alfv\'en wave or pulse.

Our results suggest that the interpretation of vortices as drivers of
Alfv\'en waves should be treated with caution. While the magnetic field
below 1~Mm may appear twisted \citep{Silva_2024a}, the SPOD analysis reveals
no coherent transverse, torsional structure in the perturbation field, as
would be expected of a torsional Alfv\'en wave
\citep{priest2014magnetohydrodynamics}. Thus, a twisted co-rotating
magnetic field likely does not, by itself, imply the excitation of a
wave or Alfv\'enic pulses, as previously assumed by
\citet{Liu_2019, Battaglia_2021}. Establishing the presence of  Alfv\'enic waves requires analysis of the perturbation field rather than the total field alone, ideally supported by diagnostics such as the Wal\'en relation.

The lack of detectable torsional Alfvénic morphology above the noise level (i.e. modes accounting for more than $0.5\%$ of the magnetic fluctuation energy) in the magnetic-energy-optimized SPOD modes below 1~Mm is consistent with torsional Alfvénic perturbations being weak or absent at these heights, with no detectable contribution to the resolved magnetic-energy fluctuations in the lower atmosphere.
Thus, the wave energy flux up to H$\alpha$ heights is likely entirely of magnetoacoustic nature, confirming the predominantly compressive nature of wave dynamics in this region of the chromosphere.
In the simulation, the waves that are present (sausage, kink/helical) do contribute significantly to the magnetic energy budget, which may explain why earlier studies, observing magnetic signatures, were tempted to interpret the structures as Alfvénic. This clearly shows that the presence of a helical mode, despite its rotational appearance, does not imply Alfv\'enic fluctuations. Such modes arise independently of magnetic field topology, as demonstrated in hydrodynamic studies \citep{dahl2021centrifugal}, and their morphological similarity to torsional Alfv\'en waves illustrates how visual identification without a dynamical test can be misleading. 

We emphasise that our findings do not preclude Alfvénic waves or pulses at greater heights. At lower plasma-$\beta$, compressive kink modes can transfer energy into localised Alfvénic fluctuations through resonant absorption; phase mixing can then progressively enhance the dissipation of this transferred energy. It is, therefore, physically plausible that the Alfvénic signatures reported at coronal heights \citep{Morton_2026} emerge from the compressive vortex-driven dynamics characterised here, although such a connection cannot be established directly from the present analysis.

Cross-validation by A-MorphIS supports the SPOD results and highlights the robustness of SPOD in detecting coherent structures in the presence of noisy, dynamic environments.
In the simulation, {SPOD analysis of $v_z$, $S_z$, and temperature shows that MHD wave activity couples kinetic, magnetic, and thermal perturbations. In observations, corresponding wave-like patterns in $v_{\mathrm{LOS}}$ and FWHM indicate that these modes drive significant vertical transport of momentum and energy.

The detection of sausage and helical wave signatures in temperature (numerical data) and FWHM (observations) indicates that the waves directly impose coherent thermal responses in the plasma. Such mode-matched imprints are unlikely to arise from passive processes and instead reflect wave-induced perturbations that actively modulate the local thermodynamic state. Across the analysed vortices, these wave modes rank among the most energetic contributors, underscoring their central role in shaping plasma motions and temperature.}

Our results demonstrate that vortex motions act as efficient waveguides in the lower atmosphere, and the Bifrost simulation reveals that they significantly amplify upward wave energy transport relative to quiescent, non-vortical regions. 
In the simulated data, the height-dependent enhancement of compressive wave energy implies that photospheric compressive waves reach the upper chromosphere mainly when guided by vortices, which act as efficient, vertically extended channels that reduce reflection and enable upward energy transport, unlike the surrounding atmosphere, where such energy is largely dissipated or trapped.
Wave energy decomposition reveals a distinct stratified regime in the vortex: compressive wave energy (linked to pressure perturbations) dominates below the H$\alpha$ formation height ($\sim$1\,Mm in our simulations), where the vertical magnetic-to-pressure energy flux ratio, $Wm_{z} / Wp_{z}$, remains less than unity. Above this height, magnetic energy becomes increasingly dominant. Nevertheless, magnetoacoustic modes continue to carry both compressive and magnetic wave energy flux throughout the lower atmosphere. 

In the Bifrost simulation, vortices produce substantial compressive wave-energy fluxes within a coronal-hole magnetic environment. At around $0.6\,\mathrm{Mm}$, the vertical compressive wave-energy flux reaches
$\sim10^4\, \mathrm{W\,m^{-2}}$ (see Appendix \ref{app:waveflux}, Fig.~\ref{fig:wave_fluxes}), overlapping with previous estimates of low-frequency acoustic-wave input into the chromosphere ($\sim2.6\,\mathrm{kW\,m^{-2}}$) in regions of predominantly vertical magnetic field \citep{Rajaguru2019}. 

The relative contribution of the compressive and magnetic components changes markedly with height. At $2\,\mathrm{Mm}$, the mean of the magnetic-to-compressive vertical-flux ratios calculated for the individual vortices, $\langle W_{m,z}/W_{p,z}\rangle$, exceeds 10 (Fig.~\ref{fig:ratios}), indicating that magnetic-energy transport increasingly dominates with height. At the same height, however, the separately averaged vertical compressive flux, $\langle W_{p,z}\rangle$, remains on the order of $\sim200\,\mathrm{W\,m^{-2}}$ (Appendix \ref{app:waveflux}, Fig.~\ref{fig:wave_fluxes}). We note that $\langle W_{m,z}/W_{p,z}\rangle$ is the mean of the individual-vortex ratios and is therefore distinct from $\langle W_{m,z}\rangle/\langle W_{p,z}\rangle$, the ratio of the separately averaged vertical fluxes shown in Fig.~\ref{fig:wave_fluxes}(b,c). In absolute energetic terms, the remaining compressive flux corresponds to approximately $14$-$20\%$ of the total coronal-hole energy requirement of $\sim1000$-$1400\,\mathrm{W\,m^{-2}}$ estimated by \citet{Withbroe1977}. Thus, although magnetic-energy transport becomes the dominant wave-energy channel at greater heights, vortex-associated compressive perturbations retain a non-negligible contribution to the upward energy flux. Moreover, the recovery of sausage and helical modes in the SPOD analysis of temperature perturbations shows that these modes are associated with thermal variability, linking the identified wave dynamics to the thermodynamic response of the chromospheric plasma.

\section*{Acknowledgements}

V.F., K.T., G.T., S.S.A.S., I.D., G.V. and I.B. are grateful to The Royal Society, International Exchanges Scheme, collaboration with Greece (IES/R1/221095). S.S.A.S., V.F. and G.V. are grateful to the Science and Technology Facilities Council (STFC) grants ST/V000977/1, ST/Y001532/1. V.F is grateful to STFC grant UKRI1165. S.J. acknowledges support from the UK Science and Technology Facilities Council (STFC) for the consolidated grants ST/T00021X/1 and ST/X000923/1 and for the small award UKRI5116. T.M.D.P. has been supported by the Research Council of Norway through its Centers of Excellence scheme, project number 262622. L.A.C.A.S., J.A.M. and G.J.J.B. acknowledge STFC for support from grant No. ST/X001008/1 and for IDL support. Computational resources have been provided by UNINETT Sigma2---the National Infrastructure for High-Performance Computing and Data Storage in Norway. V.F., K.T. and G.T. would like to thank the International Space Science Institute (ISSI) in Bern, Switzerland, for the hospitality provided to the members of the team on `The Nature and Physics of Vortex Flows in Solar Plasmas' and ‘Tracking Plasma Flows in the Sun’s Photosphere and Chromosphere: A Review \& Community Guide’.  This research has also received financial support from the European Union’s Horizon 2020 research and innovation program under grant agreement No. 824135 (SOLARNET)

\section*{Author contributions statement}
S.S.A.S conceived the study and performed the SPOD computation together with L.A.C.A.S. I.D., K.T, and G.T. prepared the observational data and performed the other wave analysis to validate the findings from SPOD. S.S.A.S, L.A.C.A.S, I.B. developed the synthetic signal. T. M. D. P. and S.J. computed the synthetic emission data from the Bifrost simulation. S.S.A.S, L.A.C.A.S, I.B., S.J., G.F., J.A.M., G.J.J.B. and V.F. worked on the identification of wave modes. S.S.A.S and I.E. performed and discussed the analysis of vortex connectivity and entropy. All authors worked on the analysis of results and reviewed the manuscript.

\textbf{Competing interests} (The authors declare no competing interests). 
\clearpage




\appendix



\section{SPOD validation}
\subsection{SPOD analysis of synthetic model \label{app:spod-model}}
\begin{figure*}
	\centering
	\includegraphics[width =0.95\textwidth]{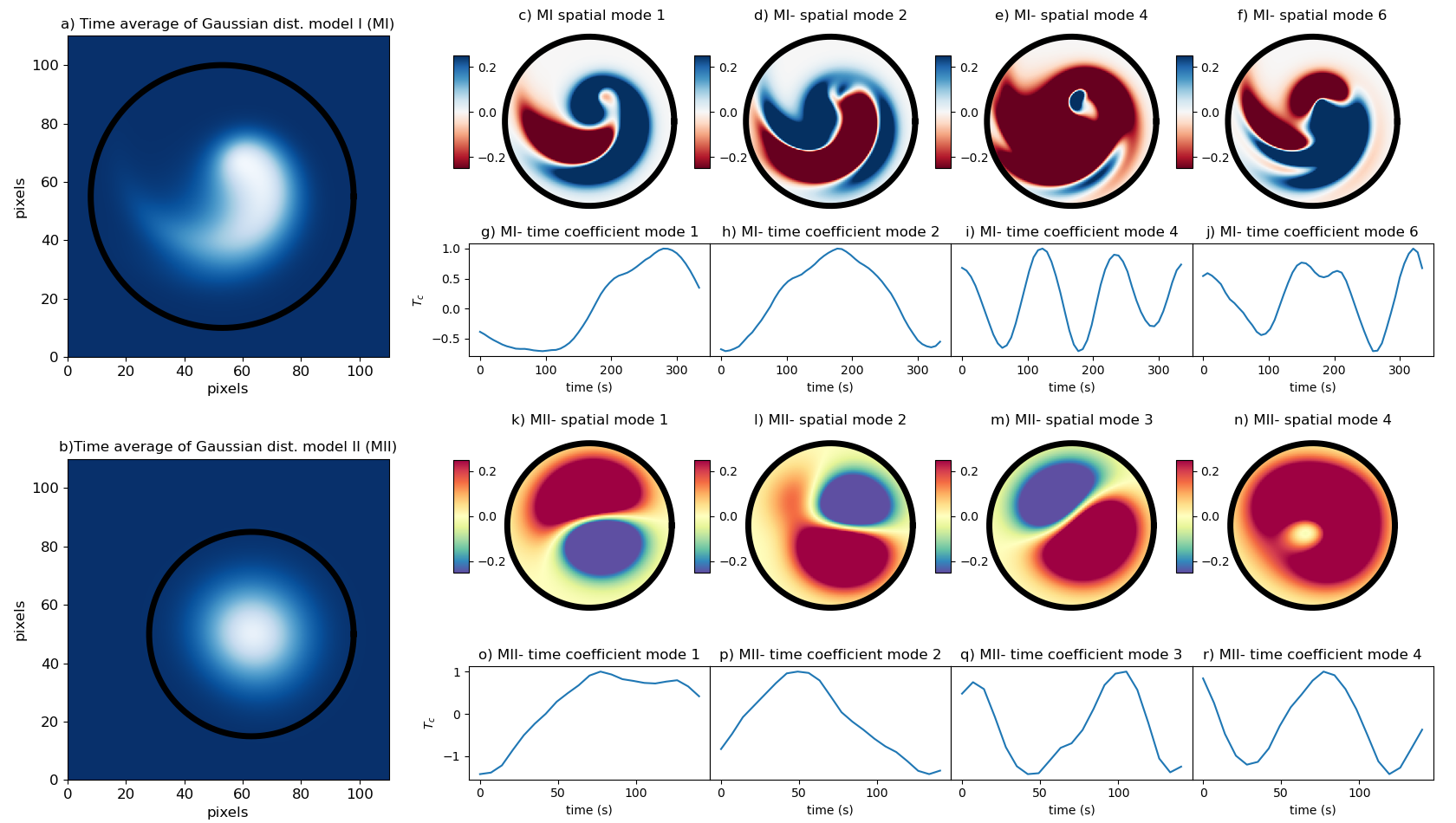}
	\caption{Synthetic vortex as waveguide modelled by a rotating Gaussian distribution. Panels (a) and (b) show the time-averaged vortex evolution regions for Models I (differential rotation) and II (rigid body rotation), with the black circle indicating the area of vortex motion. Panels (c-f) and (k-n) illustrate selected spatial SPOD modes for Models I and II, respectively, with their corresponding temporal coefficients displayed below each spatial mode. }
	\label{fig:synthetic}
\end{figure*}

The SPOD was applied to models I and II data series for selected time intervals corresponding to the average lifetime of the SST swirls and their photospheric counterparts. Both models had imposed frequencies of 9 mHz and 7 mHz for sausage and kink oscillations, respectively. In addition, we assigned a lower amplitude to the sausage modes, making the kink oscillation the dominant signal. Figure~\ref{fig:synthetic} presents the results of the synthetic data analysis. Panels (a) and (b) show the time averages for Models I and II, respectively, with the black circle marking the region of vortex evolution.  SPOD applied to the synthetic vortical waveguide shows that modes are correctly identified but their frequencies should be estimated or validated with complementary methods. Panels (c - f) and (k - n) display the selected spatial SPOD modes for models I and II, respectively, with their corresponding temporal coefficients shown below each spatial mode. Mode selection was based on the identification of patterns that resembled kink or sausage modes consistent with the imposed oscillations.  
The spatial modes reveal that the Gaussian shape, random motion, and rotation introduce significant changes to the expected wave-mode patterns. For example, in the case of solid-body rotation, the polarization direction of the kink modes is altered. In contrast, in the differential rotation case, the kink mode produces a spiral pattern (see panels (c,d) and (f)). Moreover, the sausage modes in both models resemble the spatial structure of the sausage overtone modes (panels (e) and (n)), instead of the fundamental mode imposed in the data 

To ensure physical relevance, we consider only those modes whose temporal coefficients exhibit at least quasiperiodic behaviour. Modes such as those shown in panels (c), (d), (k), and (l), whose time coefficients lack clear periodicity, are not classified as coherent wave modes.
Such differences between the imposed wave modes and the spatial modes of the SPOD may arise for different reasons. For example, a vortex flow with differential rotation that undergoes a slow body kink mode can excite helical waves. Random motion and rotation can also change the orientation of the original kink oscillatory motion, as observed in the panels (k-m). As for the sausage modes, even when removing all the other dynamics but the sausage modulation, the SPOD identifies only sausage overtones, which were not originally modelled in the synthetic data. When computing SPOD over long time intervals, we recover the original fundamental sausage mode, but there are still modes displaying sausage overtone features. This might be due to modelling the vortex as a smooth Gaussian distribution, as such profiles, when modulated, inherently excite multiple modes, including overtones, due to their spectral and geometric properties as a waveguide \citep{Gupta2024}. 

In both models, the first spatial mode corresponds to the kink oscillation, manifesting either as the kink itself or by exciting a helical wave pattern. This aligns with the imposed synthetic signal amplitudes and frequencies, where the kink mode is the dominant oscillation. However, the identification of SPOD modes as wave modes depends both on the spatial patterns and on their temporal characteristics, as shown in panels (g-j) and (o-r) of 
Fig.~\ref{fig:synthetic} for Models I and II, respectively. Due to the limited length of the time series, the leading modes exhibit time coefficients that appear wavelike but span less than one complete period. In contrast, other temporal modes display clear oscillatory behaviour with at least one full period. Consequently, modes 2 and 4 of model I and modes 3 and 4 of model II are identified as the genuine MHD wave modes. Based solely on the SPOD modes identified as wave modes, the sausage model would be dominant in the model I time series, contrary to the imposed vortex conditions. This highlights the need for caution when interpreting dominant modes in real data.

To identify the dominant frequency components of the modes, we applied the Welch method, which was employed to compute the Power Spectral Density (PSD) for their corresponding time coefficients. We obtained a relative error of  -19.01\% (model I) and 2.93\% (model II) for sausage modes. The relative errors for the kink modes were -44.65\% (models I)  and 24.28\% (models II). Thus, the frequencies of the sausage mode were recovered relatively well. For kink modes, the relative error greatly depends on the choice of the initial time and interval size used to compute the SPOD. As such, choices are limited in observational and numerical data; the synthetic data analysis shows that the mode frequencies should be estimated and/or validated by other methods.

\begin{figure*}
\centering
\includegraphics[width=0.97\linewidth]{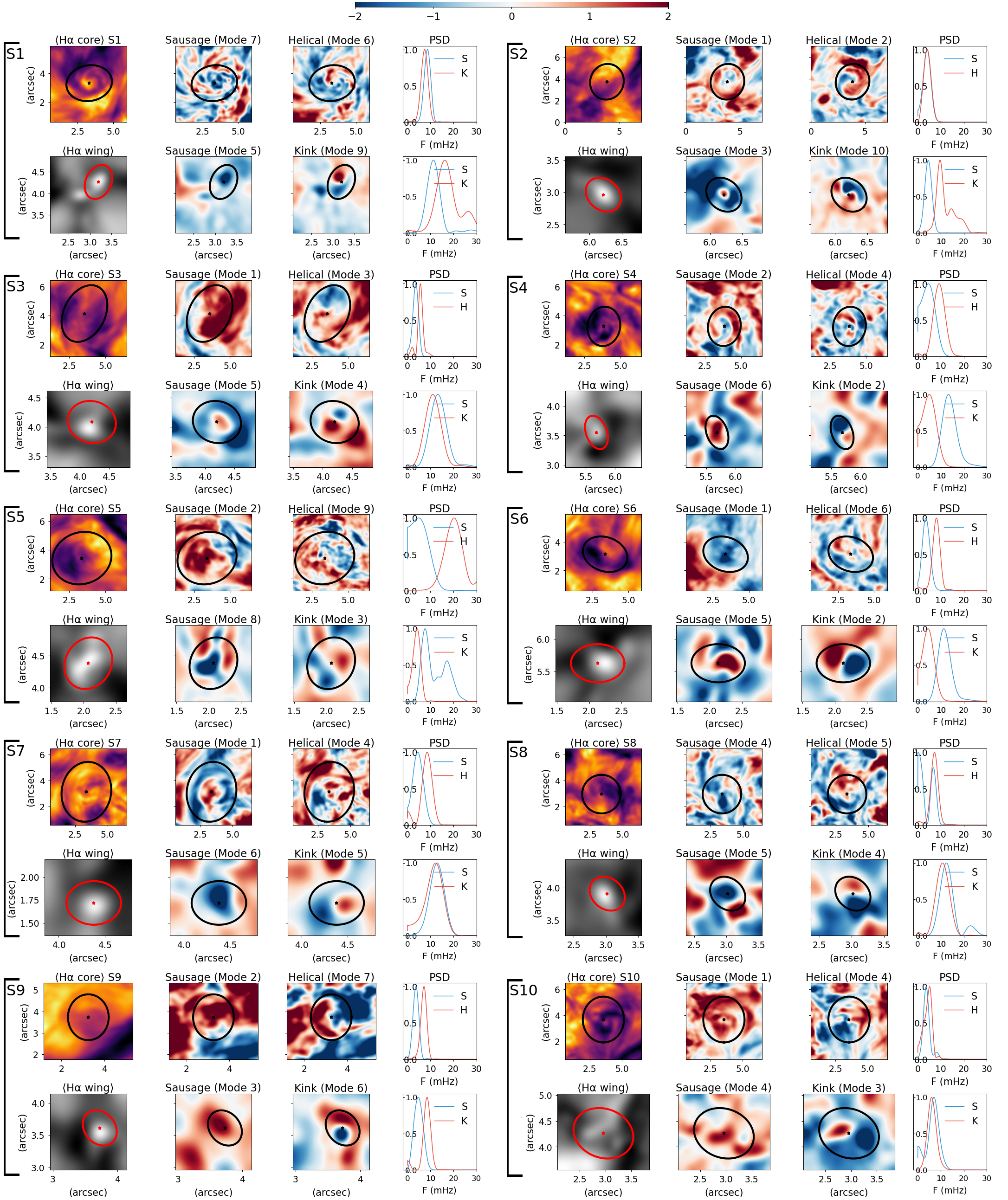}
\caption{Selected spatial modes of SPOD computed during the interval: S1 (10:21\,UT to 10:28\,UT), S2 (10:23\,UT to 10:30\,UT), S3 (10:42\,UT to 10:52\,UT), S4 (10:47\,UT to 10:51\,UT)    , S5 (10:20\,UT to 10:27\,UT), S6 (10:47\,UT to 10:53\,UT), S7 (10:41\,UT to 10:46\,UT), S8 (10:47\,UT to 10:53\,UT), S9 (10:44\,UT to 10:53\,UT), S10 (10:29\,UT to 10:37\,UT). For each swirl, the top row displays the results of H$\alpha$ at the core analysis, and the bottom row is the wing. Average H$\alpha$ core and wing emissions for the analysed time interval are shown in the first column. The black/red lines depicts the region where the vortex appears and evolves. The second column shows the identified sausage, and the third column shows the helical (top row) and kink (bottom row) modes. The last panels show the PSD of the detected temporal modes for the chromosphere and photosphere, with the red lines denoting the kink (K)/ helical (H) modes while sausage (S) modes are indicated by the blue lines.
 \label{fig:allPOD_obs}}
\end{figure*}

\begin{figure*}
\centering
\includegraphics[width=0.99\linewidth]{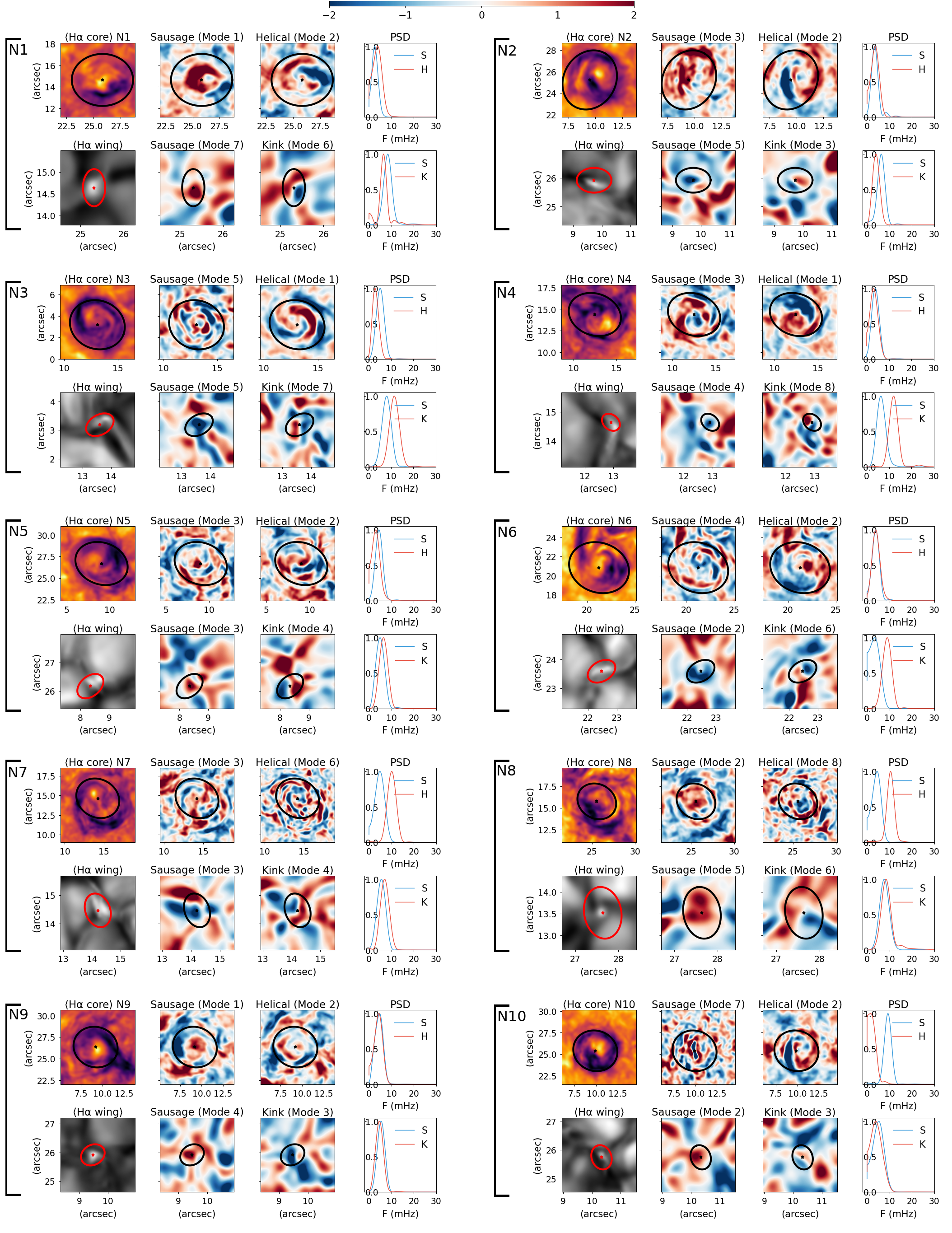}
\caption{Selected spatial modes of SPOD computed during an interval of 400 seconds within the lifetime of vortex N1-N10 for
H$\alpha$ at the core (top row) and wing (bottom row). For a description of the panels, see the caption of Fig.~\ref{fig:allPOD_obs}
 \label{fig:allPOD_num}}
\end{figure*}

The same SPOD analysis was conducted for additional selected observational vortices shown in Fig.~\ref{fig:fov}
and simulated swirls displayed in Fig.~\ref{fig:useddomains}. The obtained results are displayed in Fig.~\ref{fig:allPOD_obs} for observational data and in  Fig.~\ref{fig:allPOD_num} for Bifrost swirls. In both panels there are clear signatures of kink/helical and sausage modes.

\section{Morphological wave detection in synthetic H$\alpha$ line-center intensities \label{app:morphologic-wave}}

\begin{figure*}
\centering
    \includegraphics[width=0.95\textwidth]{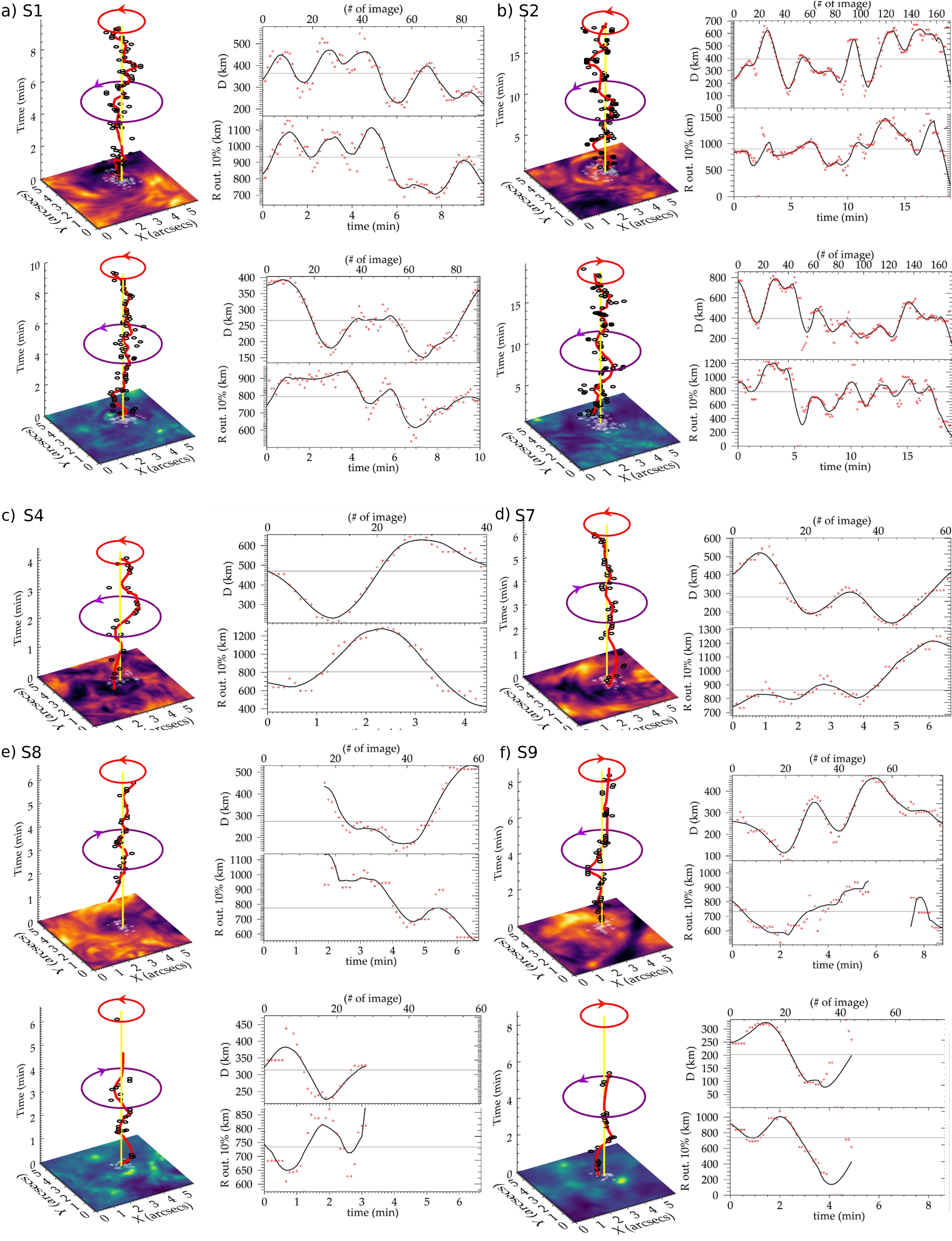} 
    \caption{Detected helical kink- and sausage-like modes in swirls S1, S2, S4, S7, S8, and S9.
    For S1 (top group): H$\alpha$ (top row) and Ca\,{\sc ii} (bottom row); S2, S8, S9, S10 follow the same two-row layout.
    S4 and S7 are shown in H$\alpha$ only.
    In the left panels, yellow vertical lines represent the detected mean centers and the red lines the curve fit of the center positions (black small circles). Red and purple circles indicate the rotation sense of the detected centers, thus of the helical kink-like mode, and of the visual vortex flow, respectively. The right panels show the temporal evolution of the instantaneous distance, $D$, of the swirl center from the mean center and of the swirl radius, $R$, derived from the outermost 10\% segments. Red symbols indicate the original data, black lines the smoothed curve fits and the horizontal grey lines the corresponding average values.}
    \label{fig:morph_all}
\end{figure*}

We now extend the analysis to additional observational vortices that are listed in Table~\ref{tab:my_label}. Using A-MorphIS, we tracked the evolution of the center and shape of each vortex to detect the signature of wave propagation. The detection algorithm was applied to chromospheric swirls observed in the H$\alpha$-0.2 \AA\ and Ca\,{\sc ii} line-center observations. For this analysis, we selected only those swirls that had minimal temporal gaps and a sufficient number of detected centers. The results, shown in Fig.~\ref{fig:morph_all}, reveal that in most cases, the observed helical wave motions align with the visual vortex flows. Moreover, they further support and cross-validate the wave propagation modes identified through the SPOD analysis, as the time evolution of the swirl center's displacement from the mean position and changes in radius suggest the presence of kink/helical-like and sausage-like modes. As summarised in Table~\ref{tab:freq}, the frequencies of these two modes tend to increase with height, since  Ca\,{\sc ii} is formed a couple of hundred kilometres lower than H$\alpha$, and are within previously reported ranges for chromospheric swirls \citep{Tziotziou_2019}. The frequency variation with height is shaped by a complex combination of factors, including magnetic field strength and geometry that also influences the local acoustic cutoff frequency, density and temperature stratification, mode conversion and reflection of waves at the magnetic canopy, as well as non-linear effects \citep{Schunker_2006,Cally_2007,Stangalini_2011,Kontogiannis_2014,Kontogiannis_2016}. Waves with frequencies lower than the chromospheric cutoff frequency can reach the chromosphere in the presence of inclined magnetic fields (ramp effect) within vortical structures \citep{Micha_1973,Suematsu_1990}.

\begin{deluxetable*}{rlll}
\digitalasset
\tablewidth{0pt}
\tablecaption{Measurements of Emission Lines” table, where f$_{\rm D}$ represents the frequency derived from the time evolution of the swirl center’s displacement from its mean position, associated with the kink/helical-like modes, and f$_{\rm R}$ the frequency derived from changes in radius, associated with sausage-like modes \label{tab:freq}}
\tablehead{
\colhead{Swirl \#} & \colhead{Spectral line} & \colhead{f$_{D}$ (mHz)} & \colhead{f$_{R}$ (mHz)}
}
\startdata
S1 &  H$\alpha$ & 3.33 & 5.56    \\
 	  & Ca\,{\sc ii}     & 3.29 & 4.4  \\
S2  &  H$\alpha$  & 2.91 & 2.9    \\
 	& Ca\,{\sc ii}  & 2.9 & 5.23  \\
S4  &  H$\alpha$  & 4.88  & 5.32  \\
S7  &  H$\alpha$ &  4.46 & 5.28  \\
S8 &  H$\alpha$ & 4.92  & 6.67  \\
 		& Ca\,{\sc ii}  & 3.23  & 6.25  \\
S9 &  H$\alpha$ & 3.75    & -  \\
 		& Ca\,{\sc ii}  & 2.5  & 2.1  \\ 
S10  &  H$\alpha$ & 2.47     &  4.15  \\
 		& Ca\,{\sc ii}  &  2.5    & 3.9 \\
\enddata
\end{deluxetable*}

\section{Wave energy flux analysis \label{app:waveflux}}
\begin{figure}
    \centering
    \includegraphics[width=0.85\linewidth]{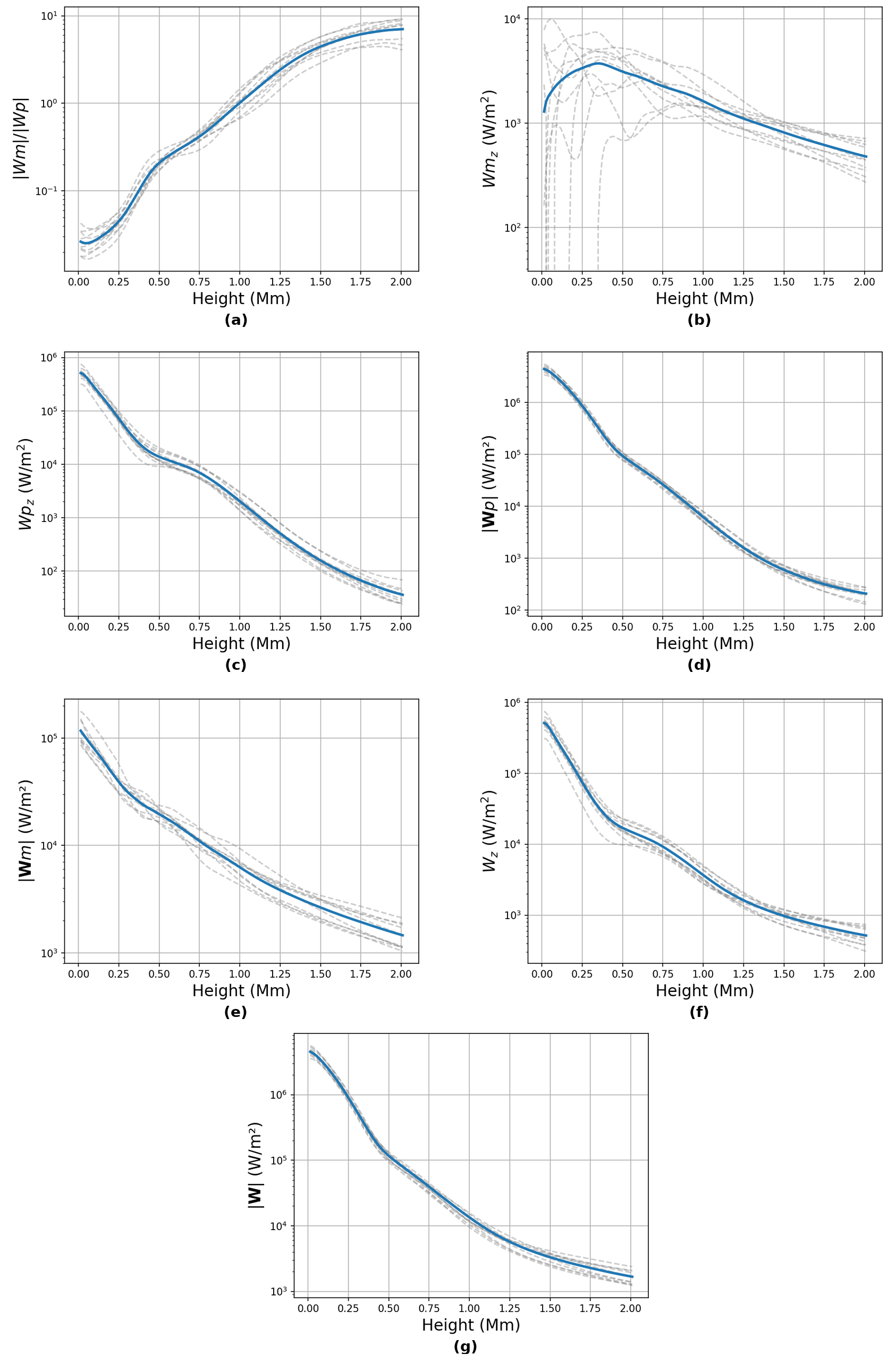}
    \caption{Wave-energy flux profiles as a function of height for the ten analysed vortices. (a) Ratio of the magnetic to compressive flux magnitudes, $|\mathbf{W}_m|/|\mathbf{W}_p|$; (b) vertical magnetic component, $W_{m,z}$; (c) vertical compressive component, $W_{p,z}$; (d) compressive flux magnitude, $|\mathbf{W}_p|$; (e) magnetic flux magnitude, $|\mathbf{W}_m|$; (f) total vertical wave-energy flux, $W_z = W_{p,z}+W_{m,z}$; and (g) magnitude of the total wave-energy flux, $|\mathbf{W}|=|\mathbf{W}_p+\mathbf{W}_m|$. The solid blue curves show the mean over all ten vortices, while the grey dashed curves show the individual vortex profiles.}
    \label{fig:wave_fluxes}
\end{figure}

To isolate the vortex region in the simulation domain, we applied a circular mask with a diameter of approximately 3.4~arcsec in the \( x\text{-}y \) plane. This relatively large radius was chosen to account for the vortex's tilt, which causes it to intersect different horizontal locations at different heights. In addition, the vortex exhibits kinking and transverse motions over time, effectively broadening the region it occupies. The chosen mask ensures that all relevant regions associated with the vortex were captured throughout its evolution. All physical quantities were spatially averaged within this disk and then temporally averaged over the same 400-second interval used in the SPOD analysis. The resulting profiles were analysed as a function of height above the photosphere and are displayed in Fig.~\ref{fig:wave_fluxes}.

\bibliography{sample701}{}

@article{dahl2021centrifugal,
  title={Centrifugal waves in tornado-like vortices: Kelvin’s solutions and their applications to multiple-vortex development and vortex breakdown},
  author={Dahl, Johannes ML},
  journal={Monthly Weather Review},
  volume={149},
  number={10},
  pages={3173--3216},
  year={2021}
}

@ARTICLE{yadav_2026,    
AUTHOR={Yadav, Nitin  and Khuman, Apanba },
TITLE={Solar vortices: catalysts of magnetoacoustic wave dissipation and atmospheric heating},         
JOURNAL={Frontiers in Astronomy and Space Sciences},       
VOLUME={Volume 13 - 2026},
YEAR={2026},
URL={https://www.frontiersin.org/journals/astronomy-and-space-sciences/articles/10.3389/fspas.2026.1819827},
DOI={10.3389/fspas.2026.1819827},
ISSN={2296-987X}}

@ARTICLE{Leenarts_2012,
       author = {{Leenaarts}, J. and {Carlsson}, M. and {Rouppe van der Voort}, L.},
        title = "{The Formation of the H{\ensuremath{\alpha}} Line in the Solar Chromosphere}",
      journal = {\apj},
         year = 2012,
        month = apr,
       volume = {749},
       number = {2},
          eid = {136},
        pages = {136},
          doi = {10.1088/0004-637X/749/2/136},
archivePrefix = {arXiv},
       eprint = {1202.1926},
 primaryClass = {astro-ph.SR},
       adsurl = {https://ui.adsabs.harvard.edu/abs/2012ApJ...749..136L}
}

@book{priest2014magnetohydrodynamics,
  title={Magnetohydrodynamics of the Sun},
  author={Priest, Eric},
  year={2014},
  publisher={Cambridge University Press}
}

@article{LUCCANEGRO2001431,
title = {Vortex breakdown: a review},
journal = {Progress in Energy and Combustion Science},
volume = {27},
number = {4},
pages = {431-481},
year = {2001},
issn = {0360-1285},
doi = {https://doi.org/10.1016/S0360-1285(00)00022-8},
url = {https://www.sciencedirect.com/science/article/pii/S0360128500000228},
author = {O Lucca-Negro and T O'Doherty}
}

@article{Albidah_2022,
doi = {10.3847/1538-4357/ac51d9},
url = {https://doi.org/10.3847/1538-4357/ac51d9},
year = {2022},
month = {mar},
publisher = {The American Astronomical Society},
volume = {927},
number = {2},
pages = {201},
author = {Albidah, A. B. and Fedun, V. and Aldhafeeri, A. A. and Ballai, I. and Brevis, W. and Jess, D. B. and Higham, J. and Stangalini, M. and Silva, S. S. A. and Verth, G.},
title = {Magnetohydrodynamic Wave Mode Identification in Circular and Elliptical Sunspot Umbrae: Evidence for High-order Modes},
journal = {The Astrophysical Journal}
}

@article{Albidah_2023,
doi = {10.3847/1538-4357/acd7eb},
url = {https://doi.org/10.3847/1538-4357/acd7eb},
year = {2023},
month = {aug},
publisher = {The American Astronomical Society},
volume = {954},
number = {1},
pages = {30},
author = {Albidah, A. B. and Fedun, V. and Aldhafeeri, A. A. and Ballai, I. and Jess, D. B. and Brevis, W. and Higham, J. and Stangalini, M. and Silva, S. S. A. and MacBride, C. D. and Verth, G.},
title = {The Temporal and Spatial Evolution of Magnetohydrodynamic Wave Modes in Sunspots},
journal = {The Astrophysical Journal}
}

@article{Alanezy2026,
author = {Alanezy, A. S. and Silva, S. S. A. and Ballai, I. and others},
title = {Oscillatory Dynamics and Energy Transport in Twisted Magnetic Flux Tubes of the Solar Photosphere},
journal = {Solar Physics},
year = {2026},
volume = {301},
pages = {61},
doi = {10.1007/s11207-026-02652-y},
url = {https://doi.org/10.1007/s11207-026-02652-y}
}

@article{Silva_2025,
doi = {10.3847/2041-8213/ae1a5e},
url = {https://doi.org/10.3847/2041-8213/ae1a5e},
year = {2025},
month = {nov},
publisher = {The American Astronomical Society},
volume = {994},
number = {1},
pages = {L27},
author = {Silva, Suzana S. A. and González-Avilés, J. J. and Riley, Pete and Ben-Nun, Michal and Rempel, Erico L. and Batista, Leonardo F. G. and Verth, Gary and Ballai, Istvan and Lin, Chia-Hsien and Schiavo, Luiz A. C. A. and Fedun, Viktor},
title = {Global Compressive Oscillations in the Outer Solar Corona Driven by an Extreme Coronal Mass Ejection},
journal = {The Astrophysical Journal Letters}
}

@article{schiavo2026waves,
author = {Schiavo, Luiz A. C. A. and Botha, Gert J. J. and McLaughlin, James A.},
title = {Wave Generation via Oscillatory Reconnection at a Three-dimensional Magnetic Null Point},
journal = {The Astrophysical Journal},
year = {2026},
volume = {999},
number = {1},
pages = {50},
doi = {10.3847/1538-4357/ae3da5}
}

@ARTICLE{graftieaux,
       author = {{Graftieaux}, Laurent and {Michard}, Marc and {Grosjean}, Nathalie},
        title = "{Combining PIV, POD and vortex identification algorithms for the study of unsteady turbulent swirling flows}",
      journal = {Measurement Science and Technology},
         year = 2001,
        month = sep,
       volume = {12},
       number = {9},
        pages = {1422-1429},
          doi = {10.1088/0957-0233/12/9/307},
       adsurl = {https://ui.adsabs.harvard.edu/abs/2001MeScT..12.1422G}
}

@ARTICLE{Morton_2026,
       author = {{Morton}, R.~J. and {Gao}, Y. and {Tajfirouze}, E. and {Tian}, H. and {Van Doorsselaere}, T. and {Schad}, T.~A.},
        title = "{Evidence for small-scale torsional Alfv{\'e}n waves in the solar corona}",
      journal = {Nature Astronomy},
         year = 2026,
        month = jan,
       volume = {10},
        pages = {42-53},
          doi = {10.1038/s41550-025-02690-9},
       adsurl = {https://ui.adsabs.harvard.edu/abs/2026NatAs..10...42M}
}

@article{Liu_2019,
  title={Evidence of ubiquitous Alfv{\'e}n pulses transporting energy from the photosphere to the upper chromosphere},
  author={Liu, Jiajia and Nelson, Chris J and Snow, Ben and Wang, Yuming and Erd{\'e}lyi, Robert},
  journal={Nature Communications},
  volume={10},
  number={1},
  pages={3504},
  year={2019},
  publisher={Nature Publishing Group UK London}
}

@article{cable_99,
author = {Cable, S. and Lin, Y.},
title = {MHD simulations of oppositely propagating Alfvén waves in the magnetosheath and solar wind},
journal = {Geophysical Research Letters},
volume = {25},
number = {11},
pages = {1821-1824},
doi = {https://doi.org/10.1029/98GL01397},
url = {https://agupubs.onlinelibrary.wiley.com/doi/abs/10.1029/98GL01397},
eprint = {https://agupubs.onlinelibrary.wiley.com/doi/pdf/10.1029/98GL01397},
year = {1998}
}

@article{ Chappell2022,
	author = {{Chappell, Bruce A.} and {Pereira, Tiago M. D.}},
	title = {SunnyNet: A neural network approach to 3D non-LTE radiative transfer},
	DOI= "10.1051/0004-6361/202142625",
	url= "https://doi.org/10.1051/0004-6361/202142625",
	journal = {Astronomy \& Astrophysics},
	year = 2022,
	volume = 658,
	pages = "A182",
}

@article{Aldhafeeri_2021,
doi = {10.3847/1538-4357/abec7a},
url = {https://dx.doi.org/10.3847/1538-4357/abec7a},
year = {2021},
month = {may},
publisher = {The American Astronomical Society},
volume = {912},
number = {1},
pages = {50},
author = {Aldhafeeri, Anwar A. and Verth, Gary and Brevis, Wernher and Jess, David B. and McMurdo, Max and Fedun, Viktor},
title = {Magnetohydrodynamic Wave Modes of Solar Magnetic Flux Tubes with an Elliptical Cross Section},
journal = {The Astrophysical Journal}
}

@article{Albidah2021,
author = {Albidah, A. B.  and Brevis, W.  and Fedun, V.  and Ballai, I.  and Jess, D. B.  and Stangalini, M.  and Higham, J.  and Verth, G. },
title = {Proper orthogonal and dynamic mode decomposition of sunspot data},
journal = {Philosophical Transactions of the Royal Society A: Mathematical, Physical and Engineering Sciences},
volume = {379},
number = {2190},
pages = {20200181},
year = {2021},
doi = {10.1098/rsta.2020.0181},

URL = {https://royalsocietypublishing.org/doi/abs/10.1098/rsta.2020.0181},
eprint = {https://royalsocietypublishing.org/doi/pdf/10.1098/rsta.2020.0181}
}

@Article{Ballai_2024,
AUTHOR = {{Ballai}, Istvan and Asiri, Fisal and Fedun, Viktor and Verth, Gary and Forgács-Dajka, Emese and Albidah, Abdulrahman B.},
TITLE = {Slow Body MHD Waves in Inhomogeneous Photospheric Waveguides},
JOURNAL = {Universe},
VOLUME = {10},
YEAR = {2024},
NUMBER = {8},
ARTICLE-NUMBER = {334},
URL = {https://www.mdpi.com/2218-1997/10/8/334},
ISSN = {2218-1997},
DOI = {10.3390/universe10080334}
}

@article{ Battaglia_2021,
	author = {Battaglia, Andrea Francesco and Canivete Cuissa, Jos\'e Roberto and Calvo, Flavio and Bossart, Aleksi Antoine and Steiner, Oskar},
	title = {The Alfv\'enic nature of chromospheric swirls},
	DOI= "10.1051/0004-6361/202040110",
	url= "https://doi.org/10.1051/0004-6361/202040110",
	journal = {Astronomy \& Astrophysics},
	year = {2021},
	volume = {649},
	pages = "A121"
}

@inproceedings{cabral_93,
author = {Cabral, Brian and Leedom, Leith Casey},
title = {Imaging vector fields using line integral convolution},
year = {1993},
isbn = {0897916018},
publisher = {Association for Computing Machinery},
address = {New York, NY, USA},
url = {https://doi.org/10.1145/166117.166151},
doi = {10.1145/166117.166151},
booktitle = {Proceedings of the 20th Annual Conference on Computer Graphics and Interactive Techniques},
pages = {263–270},
numpages = {8},
location = {Anaheim, CA},
series = {SIGGRAPH '93}
}

@ARTICLE{Dakanalis_2022,
       author = {{Dakanalis}, I. and {Tsiropoula}, G. and {Tziotziou}, K. and {Kontogiannis}, I.},
        title = "{Chromospheric swirls. I. Automated detection in H{\ensuremath{\alpha}} observations and their statistical properties}",
      journal = {Astronomy \& Astrophysics},
         year = 2022,
        month = jul,
       volume = {663},
        pages = {A94},
          doi = {10.1051/0004-6361/202243236}
}

@ARTICLE{Carlsson2016,
       author = {{Carlsson}, Mats and {Hansteen}, Viggo H. and {Gudiksen}, Boris V. and {Leenaarts}, Jorrit and {De Pontieu}, Bart},
        title = "{A publicly available simulation of an enhanced network region of the Sun}",
      journal = {Astronomy \& Astrophysics},
         year = {2016},
        month = {jan},
       volume = {585},
          eid = {A4},
        pages = {A4},
          doi = {10.1051/0004-6361/201527226},
archivePrefix = {arXiv},
       eprint = {1510.07581},
 primaryClass = {astro-ph.SR},
       adsurl = {https://ui.adsabs.harvard.edu/abs/2016A&A...585A...4C}
}

@article{Finley_2022,
	author = {{Finley, A. J.} and {Brun, A. S.} and {Carlsson, M.} and {Szydlarski, M.} and {Hansteen, V.} and {Shoda, M.}},
	title = {Stirring the base of the solar wind: On heat transfer and vortex formation},
	DOI= "10.1051/0004-6361/202243947",
	url= "https://doi.org/10.1051/0004-6361/202243947",
	journal = {A\&A},
	year = 2022,
	volume = 665,
	pages = "A118",
}

@article{Gupta2024,
  author       = {Naveen Gupta and A. K. Alex and Rohit Johari and Sanjeev Kumar},
  title        = {Excitation of frequency overtones of self-focused q-Gaussian laser beam in preformed collisional parabolic plasma channels: higher harmonic generation},
  journal      = {Journal of Optics},
  year         = {2024},
  volume       = {TBD},
  number       = {TBD},
  pages        = {TBD},
  doi          = {10.1007/s12596-024-02164-7},
  url          = {https://doi.org/10.1007/s12596-024-02164-7},
  issn         = {0974-6900}
}

@ARTICLE{Jafarzadeh_2024,
       author = {{Jafarzadeh}, S. and {Schiavo}, L.~A.~C.~A. and {Fedun}, V. and {Solanki}, S.~K. and {Stangalini}, M. and {Calchetti}, D. and {Verth}, G. and {Jess}, D.~B. and {Grant}, S.~D.~T. and {Ballai}, I. and et al.},
        title = "{Sausage, kink, and fluting magnetohydrodynamic wave modes identified in solar magnetic pores by Solar Orbiter/PHI}",
      journal = {\aap},
         year = 2024,
        month = aug,
       volume = {688},
          eid = {A2},
        pages = {A2},
          doi = {10.1051/0004-6361/202449685},
archivePrefix = {arXiv},
       eprint = {2404.18717},
 primaryClass = {astro-ph.SR},
       adsurl = {https://ui.adsabs.harvard.edu/abs/2024A&A...688A...2J}
}

@article{Maxworthy_Hopfinger_Redekopp_1985, title={Wave motions on vortex cores}, volume={151}, DOI={10.1017/S0022112085000908}, journal={Journal of Fluid Mechanics}, author={Maxworthy, T. and Hopfinger, E. J. and Redekopp, L. G.}, year={1985}, pages={141–165}}

@ARTICLE{Kuniyoshi_2023,
       author = {{Kuniyoshi}, Hidetaka and {Shoda}, Munehito and {Iijima}, Haruhisa and {Yokoyama}, Takaaki},
        title = "{Magnetic Tornado Properties: A Substantial Contribution to the Solar Coronal Heating via Efficient Energy Transfer}",
      journal = {arXiv e-prints},
         year = 2023,
        month = apr,
          eid = {arXiv:2304.03010},
        pages = {arXiv:2304.03010},
          doi = {10.48550/arXiv.2304.03010},
archivePrefix = {arXiv},
       eprint = {2304.03010},
 primaryClass = {astro-ph.SR},
       adsurl = {https://ui.adsabs.harvard.edu/abs/2023arXiv230403010K}
}

@article{Kitiashvili_2012,
doi = {10.1088/2041-8205/751/1/L21},
url = {https://dx.doi.org/10.1088/2041-8205/751/1/L21},
year = {2012},
month = {may},
publisher = {The American Astronomical Society},
volume = {751},
number = {1},
pages = {L21},
author = {I. N. Kitiashvili and A. G. Kosovichev and N. N. Mansour and A. A. Wray},
title = {DYNAMICS OF MAGNETIZED VORTEX TUBES IN THE SOLAR CHROMOSPHERE},
journal = {The Astrophysical Journal Letters}
}

@ARTICLE{Jess_2009,
       author = {{Jess}, David B. and {Mathioudakis}, Mihalis and {Erd{\'e}lyi}, Robert and {Crockett}, Philip J. and {Keenan}, Francis P. and {Christian}, Damian J.},
        title = "{Alfv{\'e}n Waves in the Lower Solar Atmosphere}",
      journal = {Science},
         year = 2009,
        month = mar,
       volume = {323},
       number = {5921},
        pages = {1582},
          doi = {10.1126/science.1168680},
archivePrefix = {arXiv},
       eprint = {0903.3546},
 primaryClass = {astro-ph.SR},
       adsurl = {https://ui.adsabs.harvard.edu/abs/2009Sci...323.1582J}
}

@book{cover2012elements,
  title={Elements of Information Theory},
  author={Cover, T.M. and Thomas, J.A.},
  isbn={9781118585771},
  lccn={2005047799},
  url={https://books.google.co.uk/books?id=VWq5GG6ycxMC},
  year={2012},
  publisher={Wiley}
}

@article{Tziotziou_2020,
	author = {{Tziotziou, K.} and {Tsiropoula, G.} and {Kontogiannis, I.}},
	title = {A persistent quiet-Sun small-scale tornado - III. Waves},
	DOI= "10.1051/0004-6361/202038951",
	url= "https://doi.org/10.1051/0004-6361/202038951",
	journal = {Astronomy \& Astrophysics},
	year = 2020,
	volume = 643,
	pages = "A166",
}

@ARTICLE{Tziotziou_2023,
       author = {{Tziotziou}, K. and {Scullion}, E. and {Shelyag}, S. and {Steiner}, O. and {Khomenko}, E. and {Tsiropoula}, G. and {Canivete Cuissa}, J.~R. and {Wedemeyer}, S. and {Kontogiannis}, I. and {Yadav}, N. and {Kitiashvili}, I.~N. and {Skirvin}, S.~J. and {Dakanalis}, I. and {Kosovichev}, A.~G. and {Fedun}, V.},
        title = "{Vortex Motions in the Solar Atmosphere}",
      journal = {Space Science Reviews},
         year = 2023,
        month = feb,
       volume = {219},
       number = {1},
          eid = {1},
        pages = {1},
          doi = {10.1007/s11214-022-00946-8},
       adsurl = {https://ui.adsabs.harvard.edu/abs/2023SSRv..219....1T}
}

@ARTICLE{Yadav_2021,
       author = {{Yadav}, N. and {Cameron}, R.~H. and {Solanki}, S.~K.},
        title = "{Vortex flow properties in simulations of solar plage region: Evidence for their role in chromospheric heating}",
      journal = {Astronomy \& Astrophysics},
         year = 2021,
        month = jan,
       volume = {645},
          eid = {A3},
        pages = {A3},
          doi = {10.1051/0004-6361/202038965},
archivePrefix = {arXiv},
       eprint = {2010.14971},
 primaryClass = {astro-ph.SR},
       adsurl = {https://ui.adsabs.harvard.edu/abs/2021A&A...645A...3Y}
}

@ARTICLE{Yadav_2020,
       author = {{Yadav}, Nitin and {Cameron}, R.~H. and {Solanki}, S.~K.},
        title = "{Simulations Show that Vortex Flows Could Heat the Chromosphere in Solar Plage}",
      journal = {The Astrophysical Journal Letters},
         year = 2020,
        month = may,
       volume = {894},
       number = {2},
          eid = {L17},
        pages = {L17},
          doi = {10.3847/2041-8213/ab8dc5},
archivePrefix = {arXiv},
       eprint = {2004.13996},
 primaryClass = {astro-ph.SR},
       adsurl = {https://ui.adsabs.harvard.edu/abs/2020ApJ...894L..17Y}
}

@article{Silva_2020,
	doi = {10.3847/1538-4357/ab99a9},
	url = {https://doi.org/10.3847/1538-4357/ab99a9},
	year = 2020,
	month = {jul},
	publisher = {American Astronomical Society},
	volume = {898},
	number = {2},
	pages = {137},
	author = {Suzana S. A. Silva and Viktor Fedun and Gary Verth and Erico L. Rempel and Sergiy Shelyag},
	title = {Solar Vortex Tubes: Vortex Dynamics in the Solar Atmosphere},
	journal = {The Astrophysical Journal}
}

@article{Silva_2021,
	doi = {10.3847/1538-4357/abfec2},
	url = {https://doi.org/10.3847/1538-4357/abfec2},
	year = 2021,
	month = {jun},
	publisher = {American Astronomical Society},
	volume = {915},
	number = {1},
	pages = {24},
	author = {Suzana S. A. Silva and Gary Verth and Erico L. Rempel and Sergiy Shelyag and Luiz A. C. A. Schiavo and Viktor Fedun},
	title = {Solar Vortex Tubes. {II}. On the Origin of Magnetic Vortices},
	journal = {The Astrophysical Journal}
}

@ARTICLE{Wedemeyer2012,
       author = {{Wedemeyer-B{\"o}hm}, Sven and {Scullion}, Eamon and {Steiner}, Oskar and
         {Rouppe van der Voort}, Luc and {de La Cruz Rodriguez}, Jaime and
         {Fedun}, Viktor and {Erd{\'e}lyi}, Robert},
        title = "{Magnetic tornadoes as energy channels into the solar corona}",
      journal = {Nature},
         year = {2012},
        month = {jun},
       volume = {486},
       number = {7404},
        pages = {505-508},
          doi = {10.1038/nature11202},
       adsurl = {https://ui.adsabs.harvard.edu/abs/2012Natur.486..505W}
}

@ARTICLE{Fedun2011,
       author = {{Fedun}, V. and {Shelyag}, S. and {Verth}, G. and {Mathioudakis}, M. and {Erd{\'e}lyi}, R.},
        title = "{MHD waves generated by high-frequency photospheric vortex motions}",
      journal = {Annales Geophysicae},
         year = {2011},
        month = {jun},
       volume = {29},
       number = {6},
        pages = {1029-1035},
          doi = {10.5194/angeo-29-1029-2011},
       adsurl = {https://ui.adsabs.harvard.edu/abs/2011AnGeo..29.1029F}
}

@ARTICLE{Giagkiozis_2018,
       author = {{Giagkiozis}, Ioannis and {Fedun}, Viktor and {Scullion}, Eamon and {Jess}, David B. and {Verth}, Gary},
        title = "{Vortex Flows in the Solar Atmosphere: Automated Identification and Statistical Analysis}",
      journal = {The Astrophysical Journal},
         year = 2018,
        month = dec,
       volume = {869},
       number = {2},
          eid = {169},
        pages = {169},
          doi = {10.3847/1538-4357/aaf797},
archivePrefix = {arXiv},
       eprint = {1706.05428},
 primaryClass = {astro-ph.SR},
       adsurl = {https://ui.adsabs.harvard.edu/abs/2018ApJ...869..169G}
}

@ARTICLE{Requerey_2017,
       author = {{Requerey}, Iker S. and {Del Toro Iniesta}, Jose Carlos and {Bellot Rubio}, Luis R. and {Mart{\'\i}nez Pillet}, Valent{\'\i}n and {Solanki}, Sami K. and {Schmidt}, Wolfgang},
        title = "{Convectively Driven Sinks and Magnetic Fields in the Quiet-Sun}",
      journal = {The Astrophysical Journal Supplement Series},
         year = 2017,
        month = mar,
       volume = {229},
       number = {1},
          eid = {14},
        pages = {14},
          doi = {10.3847/1538-4365/229/1/14},
archivePrefix = {arXiv},
       eprint = {1610.07622},
 primaryClass = {astro-ph.SR},
       adsurl = {https://ui.adsabs.harvard.edu/abs/2017ApJS..229...14R}
}

@ARTICLE{Silva_2024a,
       author = {{Silva}, Suzana and {Verth}, Gary and {Rempel}, Erico L. and {Ballai}, Istvan and {Jafarzadeh}, Shahin and {Fedun}, Viktor},
        title = "{Magnetohydrodynamic Poynting Flux Vortices in the Solar Atmosphere and Their Role in Concentrating Energy}",
      journal = {The Astrophysical Journal},
         year = 2024,
        month = mar,
       volume = {963},
       number = {1},
          eid = {10},
        pages = {10},
          doi = {10.3847/1538-4357/ad1403},
       adsurl = {https://ui.adsabs.harvard.edu/abs/2024ApJ...963...10S}
}

@ARTICLE{Silva_2024b,
       author = {{Silva}, Suzana S.~A. and {Verth}, Gary and {Ballai}, Istvan and {Rempel}, Erico L. and {Shelyag}, Sergiy and {Schiavo}, Luiz A.~C.~A. and {Gomes}, Tiago F.~P. and {Fedun}, Viktor},
        title = "{Solar Vortex Tubes. III. Vorticity and Energy Transport}",
      journal = {The Astrophysical Journal},
         year = 2024,
        month = nov,
       volume = {975},
       number = {1},
          eid = {118},
        pages = {118},
          doi = {10.3847/1538-4357/ad781a},
       adsurl = {https://ui.adsabs.harvard.edu/abs/2024ApJ...975..118S}
}

@ARTICLE{Aljohani_2022,
       author = {{Aljohani}, Yasir and {Fedun}, Viktor and {Ballai}, Istvan and {Silva}, Suzana S.~A. and {Shelyag}, Sergiy and {Verth}, Gary},
        title = "{New Approach for Analyzing Dynamical Processes on the Surface of Photospheric Vortex Tubes}",
      journal = {The Astrophysical Journal},
         year = 2022,
        month = mar,
       volume = {928},
       number = {1},
          eid = {3},
        pages = {3},
          doi = {10.3847/1538-4357/ac56db},
archivePrefix = {arXiv},
       eprint = {2202.09332},
 primaryClass = {physics.flu-dyn},
       adsurl = {https://ui.adsabs.harvard.edu/abs/2022ApJ...928....3A}
}

@ARTICLE{Tziotziou_2018,
       author = {{Tziotziou}, K. and {Tsiropoula}, G. and {Kontogiannis}, I. and {Scullion}, E. and {Doyle}, J.~G.},
        title = "{A persistent quiet-Sun small-scale tornado. I. Characteristics and dynamics}",
      journal = {Astronomy \& Astrophysics},
         year = 2018,
        month = oct,
       volume = {618},
          eid = {A51},
        pages = {A51},
          doi = {10.1051/0004-6361/201833101},
       adsurl = {https://ui.adsabs.harvard.edu/abs/2018A&A...618A..51T}
}

@ARTICLE{Tziotziou_2019,
       author = {{Tziotziou}, K. and {Tsiropoula}, G. and {Kontogiannis}, I.},
        title = "{A persistent quiet-Sun small-scale tornado. II. Oscillations}",
      journal = {Astronomy \& Astrophysics},
         year = 2019,
        month = mar,
       volume = {623},
          eid = {A160},
        pages = {A160},
          doi = {10.1051/0004-6361/201834679},
archivePrefix = {arXiv},
       eprint = {1903.04796},
 primaryClass = {astro-ph.SR},
       adsurl = {https://ui.adsabs.harvard.edu/abs/2019A&A...623A.160T}
}

@ARTICLE{Shetye_2019,
       author = {{Shetye}, Juie and {Verwichte}, Erwin and {Stangalini}, Marco and {Judge}, Philip G. and {Doyle}, J.~G. and {Arber}, Tony and {Scullion}, Eamon and {Wedemeyer}, Sven},
        title = "{Multiwavelength High-resolution Observations of Chromospheric Swirls in the Quiet Sun}",
      journal = {The Astrophysical Journal},
         year = 2019,
        month = aug,
       volume = {881},
       number = {1},
          eid = {83},
        pages = {83},
          doi = {10.3847/1538-4357/ab2bf9},
       adsurl = {https://ui.adsabs.harvard.edu/abs/2019ApJ...881...83S}
}

@ARTICLE{Maxworthy_1988,
       author = {{Maxworthy}, Tony},
        title = "{Waves on vortex cores}",
      journal = {Fluid Dynamics Research},
         year = 1988,
        month = sep,
       volume = {3},
        pages = {52-62},
          doi = {10.1016/0169-5983(88)90043-3},
       adsurl = {https://ui.adsabs.harvard.edu/abs/1988FlDyR...3...52M}
}

@ARTICLE{Kuniyoshi_2024,
       author = {{Kuniyoshi}, Hidetaka and {Bose}, Souvik and {Yokoyama}, Takaaki},
        title = "{Comprehensive Synthesis of Magnetic Tornado: Cospatial Incidence of Chromospheric Swirls and Extreme-ultraviolet Brightening}",
      journal = {The Astrophysical Journal Letters},
         year = 2024,
        month = jul,
       volume = {969},
       number = {2},
          eid = {L34},
        pages = {L34},
          doi = {10.3847/2041-8213/ad5a0e},
archivePrefix = {arXiv},
       eprint = {2404.18892},
 primaryClass = {astro-ph.SR},
       adsurl = {https://ui.adsabs.harvard.edu/abs/2024ApJ...969L..34K}
}

@ARTICLE{Skirvin_2023,
       author = {{Skirvin}, S.~J. and {Fedun}, V. and {Silva}, S.~S.~A. and {Van Doorsselaere}, T. and {Claes}, N. and {Goossens}, M. and {Verth}, G.},
        title = "{The effect of linear background rotational flows on magnetoacoustic modes of a photospheric magnetic flux tube}",
      journal = {Monthly Notices of the Royal Astronomical Society},
         year = 2023,
        month = feb,
       volume = {518},
       number = {4},
        pages = {6355-6366},
          doi = {10.1093/mnras/stac3550},
archivePrefix = {arXiv},
       eprint = {2212.00379},
 primaryClass = {astro-ph.SR},
       adsurl = {https://ui.adsabs.harvard.edu/abs/2023MNRAS.518.6355S}
}

@ARTICLE{sieber2016spectral,
       author = {{Sieber}, Moritz and {Paschereit}, C. Oliver and {Oberleithner}, Kilian},
        title = "{Spectral proper orthogonal decomposition}",
      journal = {Journal of Fluid Mechanics},
         year = 2016,
        month = apr,
       volume = {792},
        pages = {798-828},
          doi = {10.1017/jfm.2016.103},
archivePrefix = {arXiv},
       eprint = {1508.04642},
 primaryClass = {physics.flu-dyn},
       adsurl = {https://ui.adsabs.harvard.edu/abs/2016JFM...792..798S}
}

@ARTICLE{Ribeiro_2017,
       author = {{Ribeiro}, Jean H{\'e}lder Marques and {Wolf}, William Roberto},
        title = "{Identification of coherent structures in the flow past a NACA0012 airfoil via proper orthogonal decomposition}",
      journal = {Physics of Fluids},
         year = 2017,
        month = aug,
       volume = {29},
       number = {8},
          eid = {085104},
        pages = {085104},
          doi = {10.1063/1.4997202},
       adsurl = {https://ui.adsabs.harvard.edu/abs/2017PhFl...29h5104R}
}

@software{Pereira_2025,
  author       = {Tiago M. D. Pereira and
                  Edvarda Harnes},
  title        = {tiagopereira/Muspel.jl: v0.2.4},
  month        = feb,
  year         = 2025,
  publisher    = {Zenodo},
  version      = {v0.2.4},
  doi          = {10.5281/zenodo.14906962},
  url          = {https://doi.org/10.5281/zenodo.14906962},
  swhid        = {swh:1:dir:f74f0f8fb658fca7fb41d65dec2489ea28ec9e3c
                   ;origin=https://doi.org/10.5281/zenodo.10854425;vi
                   sit=swh:1:snp:b0e97c131f4b47a8aee5fea159467d406f8b
                   5649;anchor=swh:1:rel:f4f6ab1d3fbbb54b232adbad5ee1
                   6926ce2e2a25;path=tiagopereira-Muspel.jl-b857e16
                  },
}

@inproceedings{Leenaarts_2009,
	adsurl = {http://adsabs.harvard.edu/abs/2009ASPC..415...87L},
	author = {{Leenaarts}, J. and {Carlsson}, M.},
	booktitle = {The Second Hinode Science Meeting: Beyond Discovery-Toward Understanding},
	editor = {{Lites}, B. and {Cheung}, M. and {Magara}, T. and {Mariska}, J. and {Reeves}, K.},
	month = dec,
	pages = {87},
	series = {Astronomical Society of the Pacific Conference Series},
	title = {{MULTI3D: A Domain-Decomposed 3D Radiative Transfer Code}},
	volume = 415,
	year = 2009,
}

@ARTICLE{Scharmer_2008,
       author = {{Scharmer}, G.~B. and {Narayan}, G. and {Hillberg}, T. and {de la Cruz Rodriguez}, J. and {L{\"o}fdahl}, M.~G. and {Kiselman}, D. and {S{\"u}tterlin}, P. and {van Noort}, M. and {Lagg}, A.},
        title = "{CRISP Spectropolarimetric Imaging of Penumbral Fine Structure}",
      journal = {The Astrophysical Journal Letters},
         year = 2008,
        month = dec,
       volume = {689},
       number = {1},
        pages = {L69},
          doi = {10.1086/595744},
archivePrefix = {arXiv},
       eprint = {0806.1638},
 primaryClass = {astro-ph},
       adsurl = {https://ui.adsabs.harvard.edu/abs/2008ApJ...689L..69S}
}

@INPROCEEDINGS{Scharmer_2003,
       author = {{Scharmer}, Goran B. and {Bjelksjo}, Klas and {Korhonen}, Tapio K. and {Lindberg}, Bo and {Petterson}, Bertil},
        title = "{The 1-meter Swedish solar telescope}",
    booktitle = {Innovative Telescopes and Instrumentation for Solar Astrophysics},
         year = 2003,
       editor = {{Keil}, Stephen L. and {Avakyan}, Sergey V.},
       series = {Society of Photo-Optical Instrumentation Engineers (SPIE) Conference Series},
       volume = {4853},
        month = feb,
        pages = {341-350},
          doi = {10.1117/12.460377},
       adsurl = {https://ui.adsabs.harvard.edu/abs/2003SPIE.4853..341S}
}

@article{song2013,
  title={Flares and magnetic non-potentiality of NOAA AR 11158},
  author={Song, Qiao and Zhang, Jun and Yang, Shu-Hong and Liu, Yang},
  journal={Research in Astronomy and Astrophysics},
  volume={13},
  number={2},
  pages={226},
  publisher={IOP Publishing},
  year={2013},
  
}

@ARTICLE{Lofdahl_2021,
       author = {{L{\"o}fdahl}, Mats G. and {Hillberg}, Tomas and {de la Cruz Rodr{\'\i}guez}, Jaime and {Vissers}, Gregal and {Andriienko}, Oleksii and {Scharmer}, G{\"o}ran B. and {Haugan}, Stein V.~H. and {Fredvik}, Terje},
        title = "{SSTRED: Data- and metadata-processing pipeline for CHROMIS and CRISP}",
      journal = {Astronomy \& Astrophysics},
         year = 2021,
        month = sep,
       volume = {653},
          eid = {A68},
        pages = {A68},
          doi = {10.1051/0004-6361/202141326},
archivePrefix = {arXiv},
       eprint = {1804.03030},
 primaryClass = {astro-ph.IM},
       adsurl = {https://ui.adsabs.harvard.edu/abs/2021A&A...653A..68L}
}

@ARTICLE{vannoort_2005,
       author = {{Van Noort}, Michiel and {Rouppe Van Der Voort}, Luc and {L{\"o}fdahl}, Mats G.},
        title = "{Solar Image Restoration By Use Of Multi-frame Blind De-convolution With Multiple Objects And Phase Diversity}",
      journal = {Solar Physics},
         year = 2005,
        month = may,
       volume = {228},
       number = {1-2},
        pages = {191-215},
          doi = {10.1007/s11207-005-5782-z},
       adsurl = {https://ui.adsabs.harvard.edu/abs/2005SoPh..228..191V}
}

@ARTICLE{Dakanalis_2021,
       author = {{Dakanalis}, Ioannis and {Tsiropoula}, Georgia and {Tziotziou}, Kostas and {Koutroumbas}, Konstantinos},
        title = "{Automated Detection of Chromospheric Swirls Based on Their Morphological Characteristics}",
      journal = {Solar Physics},
         year = 2021,
        month = jan,
       volume = {296},
       number = {1},
          eid = {17},
        pages = {17},
          doi = {10.1007/s11207-020-01748-3},
       adsurl = {https://ui.adsabs.harvard.edu/abs/2021SoPh..296...17D}
}

@ARTICLE{Chian_2019,
       author = {{Chian}, Abraham C. -L. and {Silva}, Suzana S.~A. and {Rempel}, Erico L. and {Go{\v{s}}i{\'c}} and {}, Milan and {Bellot Rubio}, Luis R. and {Kusano}, Kanya and {Miranda}, Rodrigo A. and {Requerey}, Iker S.},
        title = "{Supergranular turbulence in the quiet Sun: Lagrangian coherent structures}",
      journal = {Monthly Notices of the Royal Astronomical Society},
         year = 2019,
        month = sep,
       volume = {488},
       number = {3},
        pages = {3076-3088},
          doi = {10.1093/mnras/stz1909},
archivePrefix = {arXiv},
       eprint = {1904.08472},
 primaryClass = {astro-ph.SR},
       adsurl = {https://ui.adsabs.harvard.edu/abs/2019MNRAS.488.3076C}
}

@ARTICLE{Kontogiannis_2014,
       author = {{Kontogiannis}, I. and {Tsiropoula}, G. and {Tziotziou}, K.},
        title = "{Transmission and conversion of magnetoacoustic waves on the magnetic canopy in a quiet Sun region}",
      journal = {Astronomy \& Astrophysics},
         year = 2014,
        month = jul,
       volume = {567},
          eid = {A62},
        pages = {A62},
          doi = {10.1051/0004-6361/201423986},
archivePrefix = {arXiv},
       eprint = {1406.5066},
 primaryClass = {astro-ph.SR},
       adsurl = {https://ui.adsabs.harvard.edu/abs/2014A&A...567A..62K}
}

@ARTICLE{Micha_1973,
       author = {{Michalitsanos}, A.~G.},
        title = "{The Five Minute Period Oscillation in Magnetically Active Regions}",
      journal = {Solar Physics},
         year = 1973,
        month = may,
       volume = {30},
       number = {1},
        pages = {47-61},
          doi = {10.1007/BF00156172},
       adsurl = {https://ui.adsabs.harvard.edu/abs/1973SoPh...30...47M}
}

@INCOLLECTION{Suematsu_1990,
       author = {{Suematsu}, Yoshinori},
        title = "{Influence of Photospheric 5-Minute Oscillations on the Formation of Chromospheric Fine Structures}",
    booktitle = {Progress of Seismology of the Sun and Stars},
         year = 1990,
       editor = {{Osaki}, Yoji and {Shibahashi}, Hiromoto},
     publisher= {Springer Berlin Heidelberg},
       volume = {367},
        pages = {211},
          doi = {10.1007/3-540-53091-6_83},
       adsurl = {https://ui.adsabs.harvard.edu/abs/1990LNP...367..211S}
}

@ARTICLE{Cally_2007,
       author = {{Cally}, P.~S.},
        title = "{What to look for in the seismology of solar active regions}",
      journal = {Astronomische Nachrichten},
         year = 2007,
        month = mar,
       volume = {328},
       number = {3},
        pages = {286},
          doi = {10.1002/asna.200610731},
       adsurl = {https://ui.adsabs.harvard.edu/abs/2007AN....328..286C}
}

@ARTICLE{Schunker_2006,
       author = {{Schunker}, H. and {Cally}, P.~S.},
        title = "{Magnetic field inclination and atmospheric oscillations above solar active regions}",
      journal = {Monthly Notices of the Royal Astronomical Society},
         year = 2006,
        month = oct,
       volume = {372},
       number = {2},
        pages = {551-564},
          doi = {10.1111/j.1365-2966.2006.10855.x},
       adsurl = {https://ui.adsabs.harvard.edu/abs/2006MNRAS.372..551S}
}

@ARTICLE{Stangalini_2011,
       author = {{Stangalini}, M. and {Del Moro}, D. and {Berrilli}, F. and {Jefferies}, S.~M.},
        title = "{MHD wave transmission in the Sun's atmosphere}",
      journal = {Astronomy \& Astrophysics},
         year = 2011,
        month = oct,
       volume = {534},
          eid = {A65},
        pages = {A65},
          doi = {10.1051/0004-6361/201117356},
archivePrefix = {arXiv},
       eprint = {1108.4576},
 primaryClass = {astro-ph.SR},
       adsurl = {https://ui.adsabs.harvard.edu/abs/2011A&A...534A..65S}
}

@ARTICLE{Kontogiannis_2016,
       author = {{Kontogiannis}, I. and {Tsiropoula}, G. and {Tziotziou}, K.},
        title = "{Wave propagation in a solar quiet region and the influence of the magnetic canopy}",
      journal = {Astronomy \& Astrophysics},
         year = 2016,
        month = jan,
       volume = {585},
          eid = {A110},
        pages = {A110},
          doi = {10.1051/0004-6361/201527053},
archivePrefix = {arXiv},
       eprint = {1511.08618},
 primaryClass = {astro-ph.SR},
       adsurl = {https://ui.adsabs.harvard.edu/abs/2016A&A...585A.110K}
}

@article{kida1981elliptical,
  title={Motion of an elliptical vortex in a uniform shear flow},
  author={Kida, Shigeo},
  journal={Journal of the Physical Society of Japan},
  volume={50},
  number={10},
  pages={3517--3520},
  year={1981},
  publisher={Physical Society of Japan}
}

@article{Rajaguru2019,
  author = {Rajaguru, S. P. and Couvidat, S. and Wachter, R.},
  title = {Low-frequency Acoustic-wave Energy Flux to the Solar Chromosphere from Hinode SOT Observations},
  journal = {The Astrophysical Journal},
  volume = {871},
  number = {2},
  pages = {155},
  year = {2019},
  doi = {10.3847/1538-4357/aaf888}
}

@article{Withbroe1977,
  author = {Withbroe, G. L. and Noyes, R. W.},
  title = {Mass and Energy Flow in the Solar Chromosphere and Corona},
  journal = {Annual Review of Astronomy and Astrophysics},
  volume = {15},
  pages = {363--387},
  year = {1977},
  doi = {10.1146/annurev.aa.15.090177.002051}
}

@ARTICLE{Mumford_2015,
       author = {{Mumford}, S.~J. and {Fedun}, V. and {Erd{\'e}lyi}, R.},
        title = "{Generation of Magnetohydrodynamic Waves in Low Solar Atmospheric Flux Tubes by Photospheric Motions}",
      journal = {The Astrophysical Journal},
         year = 2015,
        month = jan,
       volume = {799},
       number = {1},
          eid = {6},
        pages = {6},
          doi = {10.1088/0004-637X/799/1/6},
archivePrefix = {arXiv},
       eprint = {1305.7415},
 primaryClass = {astro-ph.SR},
       adsurl = {https://ui.adsabs.harvard.edu/abs/2015ApJ...799....6M}
}
\bibliographystyle{aasjournalv7}



\end{document}